\documentclass[aps,prd,floatfix,superscriptaddress,preprintnumbers]{revtex4}  
\usepackage{amssymb}
\usepackage{amsmath}
\usepackage{amsfonts}
\usepackage{epsfig}             
\usepackage{graphicx}
\usepackage{tabularx}
\usepackage{color}
\newcommand{\seq}{\begin{subequations}}
\newcommand{\sen}{\end{subequations}}
\newcommand{\eq}{\begin{eqnarray}}
\newcommand{\en}{\end{eqnarray}}
\newcommand{\la}{\langle}
\newcommand{\ra}{\rangle}

\def\shiftdown#1{#1\llap{\lower.04ex\hbox{#1}}}

\begin{document}

\title{Nucleon resonances with higher spins in soft-wall AdS/QCD} 

\author{Valery E. Lyubovitskij} 
\affiliation{Institut f\"ur Theoretische Physik,
Universit\"at T\"ubingen, 
Kepler Center for Astro and Particle Physics,  
Auf der Morgenstelle 14, D-72076 T\"ubingen, Germany}
\affiliation{Departamento de F\'\i sica y Centro Cient\'\i fico
Tecnol\'ogico de Valpara\'\i so-CCTVal, Universidad T\'ecnica
Federico Santa Mar\'\i a, Casilla 110-V, Valpara\'\i so, Chile} 
\affiliation{Department of Physics, Tomsk State University,
634050 Tomsk, Russia}
\affiliation{Tomsk Polytechnic University, 634050 Tomsk, Russia}
\author{Ivan Schmidt}
\affiliation{Departamento de F\'\i sica y Centro Cient\'\i fico
Tecnol\'ogico de Valpara\'\i so-CCTVal, Universidad T\'ecnica
Federico Santa Mar\'\i a, Casilla 110-V, Valpara\'\i so, Chile}

\date{\today}

\begin{abstract}

We present a study of  electroexcitation of nucleon 
resonances with higher spins, in a soft-wall AdS/QCD model,
comparing our results with existing data from the CLAS Collaboration 
at JLab, from MAMI, and other experiments. 

\end{abstract}

\maketitle

\section{Introduction}

Investigation of electroexcitations of nucleon resonances provides important 
information about their structure and basic properties~\cite{Devenish:1975jd,%
Capstick:1994ne,Aznauryan:2011qj,Aznauryan:2009mx,Aznauryan:2012ec,%
Mokeev:2015lda,Tiator:2011pw,Stajner:2017fmh}. 
For this reason, recent experiments 
at JLab~\cite{Aznauryan:2009mx,Aznauryan:2012ec,Mokeev:2015lda} and
at MAMI~\cite{Stajner:2017fmh,Tiator:2011pw} aim for a precise
determination of the electrocouplings of nucleon resonances and nucleons, 
supported by theoretical studies of these 
quantities~\cite{Aznauryan:2011qj,Tiator:2011pw}. 
In fact, the strong and electromagnetic structure of nucleon resonances 
has been studied in detail in many different theoretical approaches, 
such as in the MAID, SAID, and Bonn-Gatchina partial-wave analysis 
~\cite{Tiator:2011pw,Drechsel:2007if,Tiator:2008kd,Arndt:2009nv,%
Anisovich:2013jya}, 
isobar model~\cite{Aznauryan:2004jd}, constituent quark 
models~\cite{Efimov:1987sa,Bermuth:1988ms,Konen:1989jp,Stancu:1990ht,%
Li:1991yba,Lourie:1991jg,Cardarelli:1996vn,Dong:1999cz,%
Faessler:2006ky,Aznauryan:2007ja,Chen:2008as,Ramalho:2010js,%
Ramalho:2010cw,Ramalho:2011ae,Parsaei:2017pne,Ivanov:2018yul}, 
chiral approaches~\cite{Jido:2007sm,Bauer:2012at,Hilt:2017iup}, 
light-cone~\cite{Braun:2009jy} and QCD~\cite{Aliev:2013dxa}   
sum rules, light-front quark models~\cite{Capstick:1994ne,%
Obukhovsky:2011sc,Aznauryan:2012ec,Obukhovsky:2013fpa,Obukhovsky:2019aa}, 
approaches based on solutions of Bethe-Salpeter and Faddeev 
equations~\cite{Eichmann:2011vu,Wilson:2011aa,Segovia:2015hra,%
Eichmann:2016yit,Chen:2018nsg,Lu:2019bjs}, approaches used semirelativistic 
approximation and empirical parametrizations~\cite{Ramalho:2011fa,An:2008xk}, 
holographic QCD~\cite{Grigoryan:2009pp}-\cite{Gutsche:2019yoo}.    

In the past decade significant progress in the study of nucleon resonances has 
been achieved through the soft-wall AdS/QCD~\cite{deTeramond:2011qp}-\cite{Gutsche:2019yoo} formalism. 
For example, AdS/QCD is able to study the electromagnetic structure of nucleon 
and nucleon resonances in the whole region of Euclidean momentum squared $Q^2$,  
and in particular, soft-wall AdS/QCD provides the correct
power scaling description of form factors and helicity amplitudes of all hadrons
at large~$Q^2$~\cite{Brodsky:1973kr}, while it is also able to give good agreement with data
at low and intermediate $Q^2$. In Refs.~\cite{deTeramond:2011qp}-\cite{Gutsche:2019yoo} 
soft-wall AdS/QCD was focused on the study of form factors and helicity amplitudes 
of electroexcitations of the Roper $N(1440)$ (first radial excitation of the 
nucleon) and the negative-parity state $N^*(1535)$. In particular, 
in Refs.~\cite{Gutsche:2012wb,Gutsche:2017lyu,Gutsche:2019jzh,Gutsche:2019yoo,Lyubovitskij:2020otz}  
we proposed and developed a soft-wall AdS/QCD framework for the study of nucleon resonances 
with adjustable quantum numbers and successfully applied it to the unified description 
of electromagnetic structure of three states --- nucleon, Roper, and $N^*(1535)$. 
In the present manuscript we apply this theoretical approach for the study of 
the electromagnetic structure of nucleon resonances with higher spins. 

The paper is organized as follows. In Sec.~II we briefly discuss our formalism. 
In Sec.~III we present the analytical calculation and the numerical analysis 
of electromagnetic form factors and helicity amplitudes of the 
transitions between nucleon and nucleon resonances. 
Finally, Sec.~IV contains our summary.

\section{Formalism}

In this section we discuss the basic principles of our 
approach~\cite{Gutsche:2011vb}-\cite{Gutsche:2019blp} 
and focus on its application to nucleon 
resonances~\cite{Gutsche:2011vb,Gutsche:2012bp,Gutsche:2012wb,%
Gutsche:2017lyu,Gutsche:2019jzh,Gutsche:2019yoo} with higher spins. 
First, we define the conformal Poincar\'e metric, which is used in our 
formalism: 
\eq
g_{MN} \, x^M x^N = \epsilon^a_M(z) \, \epsilon^b_N(z) \, 
\eta_{ab} \, x^M x^N = \frac{1}{z^2} \, (dx_\mu dx^\mu - dz^2),
\en 
where $\epsilon^a_M(z) = \delta^a_M/z$ is the              
vielbein, $g = |{\det}(g_{MN})| = 1/z^{10}$. 

Next we discuss the construction of the effective action in terms of the 5D AdS 
fermion fields $\psi_{i,\tau}$, $\psi^{MM_1\ldots M_{l-1}}_{i,\tau}$ and 
the vector field $V_M(x,z)$, where $\tau = N + L$ is the twist, defined 
as the number of partons plus angular orbital momentum. 
The vector field is dual to the electromagnetic field, while the 
fermion fields are duals to the left- and right-handed chiral doublets of 
the nucleon and the nucleon resonances, with
${\cal O}^L = (B_1^L, B_2^L)^T$ and ${\cal O}^R = (B_1^R, B_2^R)^T$ 
where $B_1 = p, N^*_p$ and $B_2 = n, N^*_n$. These AdS fields  
are in the fundamental representations of the chiral $SU_L(2)$ 
and $SU_R(2)$ subgroups and are holographic analogs of the nucleon $N$ 
and $N^*$ resonance, respectively. 
They have constrained (confined) dynamics in AdS space, 
due to the presence of dilaton field $\varphi(z) = \kappa^2 z^2$, 
where $\kappa$ is its scale parameter. The action $S$ for the description of 
electroexcitations of nucleon resonances contains a free part $S_0$, 
describing the dynamics of AdS fields, and an interaction part $S_{\rm int}$, 
describing the interactions of fermions with the vector field dual to 
electromagnetic field   
\eq\label{actionS}
S   &=& S_0 + S_{\rm int}\,, \nonumber\\[3mm]
S_0 &=& \int d^4x dz \, \sqrt{g} \, e^{-\varphi(z)} \,
\biggl\{ {\cal L}_N(x,z) + {\cal L}_{N^*}(x,z)
+ {\cal L}_V(x,z)
\biggr\} \,, \nonumber\\[3mm]
S_{\rm int} &=& \int d^4x dz \, \sqrt{g} \, e^{-\varphi(z)} \,
{\cal L}_{VNN^*}(x,z) \,, 
\en 
where ${\cal L}_N$, ${\cal L}_{N^*}$, ${\cal L}_V$, and ${\cal L}_{VNN^*}$ 
are the free and interaction Lagrangians, respectively, given by 
\eq\label{actionS2}
{\cal L}_N(x,z) &=&  \sum\limits_{i=+,-; \,\tau} \, c_\tau \,
\bar\psi_{i,\tau}(x,z) \, \hat{\cal D}_i(z) \, \psi_{i,\tau}(x,z)
\,, \nonumber\\[3mm]
{\cal L}_{N^*}(x,z) &=&  \sum\limits_{i=+,-; \,\tau^*} \, c_{\tau^*} \,
\bar\psi^{MM_1\ldots M_{l-1}}_{i,\tau^*}(x,z) \, \hat{\cal D}_i(z) \, 
\psi_{MM_1\ldots M_{l-1},i,\tau^*}(x,z) \,, \nonumber\\[3mm]
{\cal L}_V(x,z) &=& - \frac{1}{4} V_{MN}(x,z)V^{MN}(x,z)\,, \nonumber\\[3mm]
{\cal L}_{VNN^*}(x,z) &=& \sum_{i, M} 
\sum\limits_{\tau\tau^*} \, g_{\tau\tau^*}^{(iM)} \ J_{\tau\tau^*}^{(iM)}(x,z) 
\,+\, {\rm H.c.} \,.
\en 
Here $\tau$ and $\tau^*$ are the twists of both the nucleon and nucleon resonance, 
which run from their minimal possible value. 

We have introduced the following shortened notations 
\eq
\hat{\cal D}_\pm(z) &=&  \frac{i}{2} \Gamma^M
\! \stackrel{\leftrightarrow}{\partial}_{_M} - \frac{i}{8}
\Gamma^M \omega_M^{ab} [\Gamma_a, \Gamma_b]
\, \mp \,  (\mu + U_F(z))\,, \nonumber\\
\hat{\cal V}^{N^*N}_{\pm,m}(x,z)  &=&  Q \, \Gamma^M  V_M(x,z)\,, 
\en  
$c_\tau$, $c_\tau^*$, $g_{\tau\tau^*}^{(iM)}$ and $J_{\tau\tau^*}^{(iM)}(x,z)$ 
are the sets of effective couplings and currents. 
The currents $J_{\tau\tau^*}^{(iM)}(x,z)$ with 
$i=1, 2, 3, 4$ and $M = A, B, C, D, E, F$ are given, 
in terms of AdS fermion and vector fields, by
\eq 
J_{\tau\tau^*}^{(1M)}(x,z) 
&=& J_{++,\tau\tau^*}^{(M)}(x,z) + J_{--,\tau\tau^*}^{(M)}(x,z) \,,
\nonumber\\
J_{\tau\tau^*}^{(2M)}(x,z) 
&=& J_{++,\tau\tau^*}^{(M)}(x,z) - J_{--,\tau\tau^*}^{(M)}(x,z) \,,
\nonumber\\
J_{\tau\tau^*}^{(3M)}(x,z) 
&=& J_{+-,\tau\tau^*}^{(M)}(x,z) + J_{-+,\tau\tau^*}^{(M)}(x,z) \,,
\nonumber\\
J_{\tau\tau^*}^{(4M)}(x,z) 
&=& J_{+-,\tau\tau^*}^{(M)}(x,z) - J_{-+,\tau\tau^*}^{(M)}(x,z) \,,
\en 
where 
\eq 
J_{ij,\tau\tau^*}^{(A)}(x,z) &=& \bar\psi^{MM_1\ldots M_{l-1}}_{i,\tau^*}(x,z) 
\, V^{(1)}_{MM_1 \ldots M_{l-1}}(x,z) \, \psi_{j,\tau}(x,z) \,, \nonumber\\
J_{ij,\tau\tau^*}^{(B)}(x,z) &=& \partial^K\bar\psi^{MM_1\ldots M_{l-1}}_{i,\tau^*}(x,z) 
\, V^{(2)}_{KMM_1 \ldots M_{l-1}}(x,z) \, \psi_{j,\tau}(x,z) \,, \nonumber\\
J_{ij,\tau\tau^*}^{(C)}(x,z) &=& \bar\psi^{MM_1\ldots M_{l-1}}_{i,\tau^*}(x,z) 
\, V^{(3)}_{MM_1 \ldots M_{l-1}}(x,z) \, \psi_{j,\tau}(x,z) \,, \nonumber\\
J_{ij,\tau\tau^*}^{(D)}(x,z) &=& \bar\psi^{MM_1\ldots M_{l-1}}_{i,\tau^*}(x,z) \, 
i\Gamma^z \, V^{(1)}_{MM_1 \ldots M_{l-1}}(x,z) \, \psi_{j,\tau}(x,z) \,, \nonumber\\
J_{ij,\tau\tau^*}^{(E)}(x,z) &=& \partial^K\bar\psi^{MM_1\ldots 
M_{l-1}}_{i,\tau^*}(x,z) \, i\Gamma^z 
\, V^{(2)}_{KMM_1 \ldots M_{l-1}}(x,z) \, \psi_{j,\tau}(x,z) \,, \nonumber\\
J_{ij,\tau\tau^*}^{(F)}(x,z) &=& \bar\psi^{MM_1\ldots M_{l-1}}_{i,\tau}(x,z) \, 
i\Gamma^z \, V^{(3)}_{MM_1 \ldots M_{l-1}}(x,z) \, \psi_{j,\tau}(x,z) \nonumber\\
\en 
and 
\eq
V^{(1)}_{MM_1 \ldots M_{l-1}}(x,z) &=& 
\partial_{M_1} \ldots \partial_{M_{l-1}} \, 
i \Gamma^K V_{KM}(x,z) \,, \nonumber\\ 
V^{(2)}_{KMM_1 \ldots M_{l-1}}(x,z) &=& 
\partial_{M_1} \ldots \partial_{M_{l-1}} \, 
V_{KM}(x,z) \,, \nonumber\\
V^{(3)}_{MM_1 \ldots M_{l-1}}(x,z) &=& 
\partial_{M_1} \ldots \partial_{M_{l-1}} \,
\partial^K V_{KM}(x,z) \,, 
\en
Here 
$\mu$ is the five-dimensional mass of the spin-$\frac{1}{2}$ AdS      
fermion with $\mu = 3/2 + L$ ($L$ is the orbital angular momentum);      
$U_F(z) = \varphi(z)$ is the dilaton potential; 
$Q = {\rm diag}(1,0)$ is the charge matrix corresponding to doublets of 
nucleon (nucleon resonances); 
$V_{MN} = \partial_M V_N - \partial_N V_M$ is                            
the stress tensor for the vector field;  
$\omega_M^{ab} = (\delta^a_M \delta^b_z - \delta^b_M \delta^a_z)/z$ is 
the spin connection term; while $\Gamma^M = \epsilon^M_a \Gamma^a$ and 
$\Gamma^a = (\gamma^\mu, -i \gamma^5)$ are the Dirac matrices in AdS space, 
$[\Gamma_a, \Gamma_b] = \Gamma_a \Gamma_b - \Gamma_b \Gamma_a$.

Next we split 5D AdS fermion fields $\psi_{\pm,\tau}(x,z)$ and
$\psi^{MM_1\ldots M_{l-1}}_{\pm,\tau}(x,z)$ into left- and right-chirality components 
\eq\label{psi_expansion_Nucleon}
\psi(x,z) = \psi_L(x,z) + \psi_R(x,z)\,, \quad 
\psi_{L/R}(x,z) = \frac{1 \mp \gamma^5}{2} \, \psi(x,z)
\en
for the nucleon, and
\eq\label{psi_expansion_Nstar} 
\psi^{MM_1\ldots M_{l-1}}(x,z) = 
\psi_L^{MM_1\ldots M_{l-1}}(x,z) + \psi_R^{MM_1\ldots M_{l-1}}(x,z)\,, \quad 
\psi_{L/R}^{MM_1\ldots M_{l-1}}(x,z)
= \frac{1 \mp \gamma^5}{2} \, \psi^{MM_1\ldots M_{l-1}}(x,z )
\en
for the nucleon resonances  with higher spins 
and perform the Kaluza-Klein expansion as:
\eq 
\psi_{\pm,\tau}(x,z) &=& \frac{1}{\sqrt{2}} \, \sum\limits_n \, 
\biggl[
\pm \psi_{L n}(x) \ F^{L/R}_{\tau n}(z)
+   \psi_{R n}(x) \ F^{R/L}_{\tau n}(z)\biggr]\,, \\
\psi_{\pm,\tau}^{MM_1\ldots M_{l-1}}(x,z) &=&
 \frac{1}{\sqrt{2}} \, \sum\limits_n \, \epsilon^{MM_1\ldots M_{l-1}}_{aa_1\ldots a_{l-1}}(z)\, 
\biggl[
\pm \psi_{L n}^{aa_1\ldots a_{l-1}}(x) \ F^{L/R}_{\tau n}(z) 
+   \psi_{R n}^{aa_1\ldots a_{l-1}}(x) \ F^{R/L}_{\tau n}(z)\biggr]\,,
\en 
where $n$ is the radial quantum number and 
\eq
\epsilon^{MM_1\ldots M_{l-1}}_{aa_1\ldots a_{l-1}}(z) =
\epsilon^{M}_{a}(z)\,\epsilon^{M_1}_{a_1}(z) \ldots
\epsilon^{M_{l-1}}_{a_{l-1}}(z) \,.
\en
Here 
\eq\label{psi_expansion1}
F^{L/R}_{\tau n}(z) = e^{\kappa^2 z^2/2} \, z^2 \, f^{L/R}_{\tau n}(z),
\en 
are the bulk profiles with twist $\tau$ and radial quantum number $n$, 
which depend on the holographic variable $z$, where  
\eq\label{fL_fR} 
f^L_{\tau n}(z) &=& \sqrt{\frac{2 \Gamma(n+1)}{\Gamma(\tau+n)}} \, \kappa^{\tau} \,
z^{\tau - 1/2} \, e^{-\kappa^2 z^2/2} \, L_n^{\tau-1}(\kappa^2 z^2)\,, \nonumber\\
f^R_{\tau n}(z) &=& \sqrt{\frac{2 \Gamma(n+1)}{\Gamma(\tau-1+n)}}  \, \kappa^{\tau-1} \,
z^{\tau - 3/2} \, e^{-\kappa^2 z^2/2} \, L_n^{\tau-2}(\kappa^2 z^2) 
\en 
and $L_n^m(x)$ are the generalized Laguerre polynomials. 
The bulk profiles $f^{L/R}_{\tau}(z)$ are normalized as 
\eq 
1 = \int\limits_0^\infty dz \, \Big[f^L_{\tau n}(z)\Big]^2 
  = \int\limits_0^\infty dz \, \Big[f^R_{\tau n}(z)\Big]^2 \,.
\en 
The nucleon is identified as the ground state with $n=L=0$, while 
the nucleon resonance have specific values of $n$ and $L$. 
In Table~\ref{tab:baryons} we display the quantum numbers 
(spin-parity $J^P$, angular orbital moment $L$,  
radial quantum number $n$, mass) of the baryons considered 
in the present paper. The action describing transitions 
$\frac{1}{2}^+ \gamma^* \to \frac{1}{2}^\pm$ 
has been derived and discussed in detail 
in Refs.~\cite{Gutsche:2012wb,Gutsche:2017lyu,Gutsche:2019jzh,Gutsche:2019yoo}. 
In Appendix~\ref{action12} we briefly specify this action.

\begin{table}[htb]
\begin{center}
\caption{Quantum numbers of nucleon and nucleon resonances} 
\vspace*{.1cm}

\def\arraystretch{1.25}
\begin{tabular}{|c|c|c|c|c|}
\hline
 Baryon & $J^P$ & $L$ & $n$ & Mass (MeV)~\cite{PDG20} \\
\hline
$N(938)$  & $\frac{1}{2}^+$ & 0 & 0 & 938.27 \\
\hline
$N(1440)$ & $\frac{1}{2}^+$ & 0 & 1 & 1370 $\pm$ 10 \\  
\hline
$N(1710)$ & $\frac{1}{2}^+$ & 0 & 2 & 1700 $\pm$ 20 \\
\hline
$N(1535)$ & $\frac{1}{2}^-$ & 1 & 0 & 1510 $\pm$ 10 \\ 
\hline
$N(1520)$ & $\frac{3}{2}^-$ & 1 & 0 & 1510 $\pm$  5 \\
\hline
$N(1650)$ & $\frac{1}{2}^-$ & 1 & 0 & 1655 $\pm$ 15 \\ 
\hline
$N(1700)$ & $\frac{3}{2}^-$ & 1 & 0 & 1700 $\pm$ 50 \\ 
\hline
$N(1675)$ & $\frac{5}{2}^-$ & 1 & 0 & 1660 $\pm$ 5  \\
\hline
$N(1720)$ & $\frac{3}{2}^+$ & 2 & 0 & 1675 $\pm$ 15 \\ 
\hline
$N'(1720)$ & $\frac{3}{2}^+$ & 2 & 0 & 1720 $\pm$ 15~\cite{Mokeev:2020hhu} \\  
\hline
$N(1680)$ & $\frac{5}{2}^+$ & 2 & 0 & $1675^{+5}_{-10}$ \\
\hline
$\Delta(1232)$ & $\frac{3}{2}^+$ & 0 & 0 & 1210 $\pm$ 1 \\
\hline
$\Delta(1620)$ & $\frac{1}{2}^-$ & 1 & 0 & 1600 $\pm$ 10 \\
\hline
$\Delta(1700)$ & $\frac{3}{2}^-$ & 1 & 0 & 1665 $\pm$ 25 \\
\hline
\end{tabular}
\label{tab:baryons}
\end{center}
\end{table}

For the vector field $V_\mu(x,z)$ we apply the axial gauge 
$V_z = 0$ and perform a Fourier transformation 
with respect to the Minkowski coordinate
\eq\label{V_Fourier}
V_\mu(x,z) = \int \frac{d^4q}{(2\pi)^4} e^{iqx} V_\mu(q) V(q,z)\,, 
\en
where $V(q,z)$ is the vector bulk-to-boundary (dual to the 
$q^2$-dependent electromagnetic current) obeying the equation 
of motion 
\eq
\partial_z \biggl( \frac{e^{-\varphi(z)}}{z} \,
\partial_z V(q,z)\biggr) + q^2 \frac{e^{-\varphi(z)}}{z} \, 
V(q,z) = 0 
\en 
with solution in terms of gamma $\Gamma(n)$ and 
Tricomi $U(a,b,z)$ functions 
\eq
\label{VInt_q}
V(q,z) = \Gamma\Big(1 - \frac{q^2}{4\kappa^2}\Big)
\, U\Big(-\frac{q^2}{4\kappa^2},0,\kappa^2 z^2\Big) \,.
\en
In was shown in Ref.~\cite{Grigoryan:2007my} 
that in the Euclidean region ($Q^2 = - q^2 > 0$)
it is convenient to use the integral
representation for $V(Q,z)$
\eq
\label{VInt}
V(Q,z) = \kappa^2 z^2 \int_0^1 \frac{dx}{(1-x)^2}
\, x^{a} \,
e^{- \kappa^2 z^2 \frac{x}{1-x} }\,, \qquad 
a = Q^2/(4 \kappa^2)\,. 
\en

The sets of parameters 
$c_\tau$, $c_{\tau^*}$, and $g_{\tau\tau^*}^{(iM)}$ 
induce mixing of the contributions of AdS 
fields with different twist dimensions. 
The parameters $c_\tau$ and $c_{\tau^*}$ 
are constrained by the conditions 
$\sum_\tau \, c_\tau = 1$ 
and $\sum_\tau \, c_{\tau^*} = 1$, 
to guarantee the correct normalization of the 
kinetic terms $\bar\psi(x)i\!\!\not\!\partial\psi(x)$
of the four-dimensional spinor fields. 
This condition is also consistent with 
electromagnetic gauge invariance 
(see details in Refs.~\cite{Gutsche:2012bp,Gutsche:2012wb}).   
Therefore, the masses of the nucleon and nucleon resonance 
are identified by the expressions~\cite{Gutsche:2012bp,Gutsche:2012wb}
\eq\label{Matching1}
M_N = 2 \kappa \sum\limits_\tau\, c_\tau\, \sqrt{\tau - 1}
\,, \quad\quad  
M_{N^*} &=& 2 \kappa \sum\limits_\tau \, c_{\tau}^*\, \sqrt{\tau-1}\,, 
\en 
where the leading twist from which the sums start in 
Eq.~(\ref{Matching1}) is defined as $\tau = 3 + L$, where $L$ is 
the angular orbital moment specified for baryons 
in Table~\ref{tab:baryons}. 

The baryon form factors are determined analytically
using the bulk profiles of fermion fields
and the bulk-to-boundary propagator $V(Q,z)$ of 
the vector field (for exact expressions see the next section). 
The calculational technique was already described in detail
in Refs.~\cite{Gutsche:2012bp,Gutsche:2012wb,Gutsche:2017lyu}.
The parameter $\kappa = 383$ MeV is universal and was fixed 
in previous studies (see, e.g., Refs.~\cite{Gutsche:2012bp,Gutsche:2012wb}), while
the other parameters are fixed from a fit to the helicity amplitudes of 
the $\gamma^* N \to N^*$ transitions. 
   
\section{Electromagnetic form factors and helicity 
amplitudes of the $\gamma^* N \to N^*$ transitions}

Due to Lorenz covariance and gauge invariance, the matrix elements of 
the electromagnetic $\gamma^* N \to N^*$ transitions 
can be expressed in terms of their general
Lorenz structures as
\eq 
H_{\mu\nu}^{(1)} = \not\! q g_{\mu\nu} - \gamma_\mu q_\nu\,, \quad 
H_{\mu\nu}^{(2)} = p_{1\mu} q_\nu - g_{\mu\nu} \, p_1q\,, \quad 
H_{\mu\nu}^{(3)} = q_\mu q_\nu - g_{\mu\nu} q^2 
\en 
and the relativistic form factors $G_{i}(Q^2)$, $i=1,2,3$ 
as~\cite{Devenish:1975jd,Capstick:1994ne,Aznauryan:2011qj,%
Kadeer:2005aq,Faessler:2009xn}   
\eq\label{matrix_elements}
\la N^* |J_\mu^{\rm em}|N \ra &=& 
 \bar u_{N^*}^{\nu\nu_1 \ldots \nu_{l-1}}(p_1\lambda_1)
\, q_{\nu_1} \ldots q_{\nu_{l-1}} \, 
\biggl(
\begin{array}{c}
-\gamma_5 \\
I         \\
\end{array}
\biggr) \, \Gamma_{\mu\nu}(Q^2) \, u_N(p_2\lambda_2) \,, \nonumber\\ 
\Gamma_{\mu\nu}(Q^2) &=& 
G_1(Q^2) \, H_{\mu\nu}^{(1)} \,+\, 
G_2(Q^2) \, H_{\mu\nu}^{(2)} \,+\, 
G_3(Q^2) \, H_{\mu\nu}^{(3)} 
\,, 
\en
Here $u_{N}(p_2\lambda_2)$ and 
$u_{N^*}^{\nu\nu_1 \ldots \nu_{l-1}}(p_1\lambda_1)$ are 
spin-$\frac{1}{2}$ and higher spin (Rarita-Schwinger) spinors, 
respectively. The Rarita-Schwinger spinor satisfies the conditions 
\eq 
\bar u_{N^*}^{\nu\nu_1 \ldots \nu_{l-1}} \, \gamma_{\alpha} =  
\bar u_{N^*}^{\nu\nu_1 \ldots \nu_{l-1}} \, q_{\alpha} = 0 \qquad 
{\rm for} \  \alpha \in \{\nu,\nu_1,\ldots,\nu_{l-1}\}
\,,
\en 
$q = p_1 - p_2$, and $\lambda_1$, $\lambda_2$, and $\lambda$ are 
the helicities of the final, initial baryon and photon, respectively, 
with the relation $\lambda_2 = \lambda_1 - \lambda$. 
In the rest frame of the $N^*$ the four momenta of $N^*$, $N$,
photon and the polarization vector of photon are specified as:
\eq
& &p_1 = (M_1, \vec{0\,})\,, \quad 
   p_2 = (E_2, 0, 0, -|{\bf p}|)\,, \quad  
     q = (q^0, 0, 0,  |{\bf p}|)\,, \nonumber\\
& &\epsilon^\mu(\pm) = (0, - \vec{\epsilon\,}^{\pm})\,, \quad 
\vec{\epsilon\,}(\pm) = \frac{1}{\sqrt{2}} (\pm 1, i, 0)\,, \quad 
\epsilon^\mu(0) = \frac{1}{\sqrt{Q^2}}(|{\bf p}|, 0, 0, q^0)\,,  
\en
where
$ |{\bf p}| = \frac{\sqrt{Q_+ Q_-}}{2M_1}$ 
is the absolute value of the three-momentum of the nucleon or the photon, 
$Q_\pm = M_\pm^2 + Q^2$, and $M_\pm = M_1 \pm  M_2$.

It is convenient to introduce the helicity amplitudes 
$A_{1/2}$, $A_{3/2}$, and $S_{1/2}$ responsible for the 
helicity transitions 
$\lambda_2 = \pm\frac{1}{2} \to \lambda_1 = \mp\frac{1}{2}$, 
$\lambda_2 = \pm\frac{1}{2} \to \lambda_1 = \pm\frac{3}{2}$, and 
$\lambda_2 = \pm\frac{1}{2} \to \lambda_1 = \pm\frac{1}{2}$, 
respectively, which are related to the invariant form factors $G_i(Q^2)$ 
as~\cite{Devenish:1975jd,Capstick:1994ne,Aznauryan:2011qj}    
\eq 
A_{1/2}(Q^2) = b \, h_3(Q^2)\,, \quad 
A_{3/2}(Q^2) = \pm \frac{b \, \sqrt{3}}{l} \, h_2(Q^2) \,, \quad 
S_{1/2}(Q^2) &=& \frac{b \, |{\bf p}|}{M_1 \sqrt{2}} \, h_1(Q^2)\,, 
\en 
where 
\eq
h_1(Q^2) &=& \pm 4 M_1 G_1(Q^2) + 4  M_1^2 G_2(Q^2)
+ 2 (M_+ M_- - Q^2) G_3(Q^2) \,, \nonumber\\
h_2(Q^2) &=& \mp 2 M_\pm G_1(Q^2)
- (M_+ M_- - Q^2) G_2(Q^2) + 2Q^2 G_3(Q^2)\,, \nonumber\\
h_3(Q^2) &=& - \frac{2}{M_1} \, (M_\pm M_2 \pm Q^2) G_1(Q^2)
+ (M_+ M_- - Q^2) G_2(Q^2) - 2Q^2 G_3(Q^2)\,,
\en
\eq
b = |{\bf p}|^{l-1} \, \sqrt{\frac{\pi \alpha Q_\mp}
{8 M_+ M_- M_2 c_{l+1}}}\,, \quad
c_l = \frac{(2l)!}{2^l (l!)^2} \,, \quad l = J - \frac{1}{2}
\,,
\en
where $\alpha = 1/137.036$ is the fine-structure constant. 

The relations expressing the relativistic form factors $G_i$ through 
the set of form factors $h_i$ and helicity amplitudes read: 
\eq
G_1(Q^2) &=& \mp \frac{h_2(Q^2)+h_3(Q^2)}{2 Q_+} \, M_1
= \mp \frac{M_1}{2 Q_\pm b} \,
\biggl[A_{1/2}(Q^2) \pm \frac{l}{\sqrt{3}} \, A_{3/2}(Q^2)
\biggr] \,,
\nonumber\\
G_2(Q^2) &=& \frac{1}{Q_+ Q_-} \, \biggl[
h_1(Q^2) Q^2 + h_2(Q^2) (Q^2 \mp M_\mp M_2) + h_3(Q^2) M_\mp  M_1 \biggr]
\nonumber\\
&=&
\frac{1}{Q_+ Q_- b} \, \biggl[ S_{1/2}(Q^2) \frac{Q^2 M_1 \sqrt{2}}{|{\bf p}|}
\pm \frac{l}{\sqrt{3}} \, A_{3/2}(Q^2) \, (Q^2 \mp M_\mp M_2) +
A_{1/2}(Q^2) M_\mp M_1 \biggr] \,,
\\
G_3(Q^2) &=& \frac{1}{Q_+ Q_-} \, \biggl[
h_1(Q^2) (M_+ M_- -Q^2) + (h_2(Q^2) - h_3(Q^2)) M_1^2
\biggr]
\nonumber\\
&=&
\frac{1}{Q_+ Q_- b} \, \biggl[ S_{1/2}(Q^2) \frac{M_1}{|{\bf p}| \sqrt{2}}
(M_+M_- - Q^2) + \biggl(\pm \frac{l}{\sqrt{3}} \, A_{3/2}(Q^2) -
A_{1/2}(Q^2) \biggr) M_1^2 \biggr] \,, \nonumber
\en

The structure of the $\frac{1}{2}^+ \gamma^* \to \frac{1}{2}^\pm$ 
is simpler and is given by 
the form 
\eq\label{matrix_elements_one_half}
M^\mu(p_1\lambda_1,p_2\lambda_2) &=& \bar u_{N^*}(p_1\lambda_1)
\biggl[ \gamma^\mu_\perp \, F_1^{N^*N}(-q^2)
+ i \sigma^{\mu\nu} \frac{q_\nu}{M_+} \, F_2^{N^*N}(-q^2)
\, \biggr] \, \biggl(
\begin{array}{c}
I        \\
\gamma_5 \\
\end{array}
\biggr)
\, u_{N}(p_2\lambda_2) \,,  
\en 
where $\gamma^\mu_\perp = \gamma^\mu - q^\mu \not\! q/q^2$. 

The helicity amplitudes defining the $\frac{1}{2}^+ \to \frac{1}{2}^+$
and $\frac{1}{2}^+ \to \frac{1}{2}^-$ transitions in terms of form factors 
are defined, respectively, as: 
\eq
A_{1/2}^\pm(Q^2) &=& \sqrt{\frac{2 \pi \alpha Q_\mp}{M_1 M_2 E}}  \,
\biggl[ F_1^{N^*N}(Q^2) + F_2^{N^*N}(Q^2) \frac{M_\pm}{M_+} \biggr] \,,
\nonumber\\
S_{1/2}^\pm(Q^2) &=& \pm \frac{|{\bf p}|}{M_+} \,
\sqrt{\frac{\pi \alpha Q_\mp}{M_1 M_2 E}} \,
\biggl[ F_1^{N^*N}(Q^2) \frac{M_\pm M_+}{Q^2} - F_2^{N^*N}(Q^2)  \biggr] \,.
\en
 
In the case of the high-spin resonances, the set of helicity amplitudes $(A_{1/2}, A_{3/2}, S_{1/2})$ is related 
with the set of the charge $(G_E)$, magnetic $(G_M)$, and Coulomb 
$(G_Q)$ form factors~\cite{Devenish:1975jd,Capstick:1994ne,Aznauryan:2011qj}:  
\eq
G_E &=& - F_l^+ \, \frac{2}{l+1} \,
\biggl( \frac{l}{\sqrt{3}} \, A_{3/2} \,-\, A_{1/2} \biggr) \,,
\nonumber\\
G_M &=& - F_l^+ \, \frac{2l}{l+1} \,
\biggl( \frac{l+2}{\sqrt{3}} \, A_{3/2} \,+\, A_{1/2} \biggr) \,,
\\
G_C &=& 2 \sqrt{2} F_l^+ \, \frac{M_1}{|{\bf p}|} \,
S_{1/2} \nonumber
\en
for abnormal parity transitions
$\frac{1}{2}^+ \to \frac{3}{2}^+, \frac{5}{2}^-, \ldots$
\eq
G_E &=& - F_l^- \, \frac{2l}{l+1} \,
\biggl( \frac{l+2}{\sqrt{3}} \, A_{3/2} \,+\, A_{1/2} \biggr) \,,
\nonumber\\
G_M &=& - F_l^- \, \frac{2}{l+1} \,
\biggl( \frac{l}{\sqrt{3}} \, A_{3/2} \,-\, A_{1/2} \biggr) \,,
\\
G_C &=& 2 \sqrt{2} F_l^- \, \frac{M_1}{|{\bf p}|} \,
S_{1/2} \nonumber
\en
for normal parity transitions
$\frac{1}{2}^+ \to \frac{3}{2}^-, \frac{5}{2}^+, \ldots$

In terms of the relativistic form factor $h_i$, the  
charge, magnetic, and Coulomb form factors are  
expressed as 
\eq
G_E &=& - \frac{2b F_l^+}{l+1} \, (h_2-h_3)\,,
\nonumber\\
G_M &=& - \frac{2b F_l^+}{l+1} \, \Big[ l (h_2+h_3) + 2h_2 \Big] \,,
\\
G_E &=& 2b F_l^+ \, h_1 \nonumber
\en
for abnormal parity transitions and
\eq
G_E &=& \frac{2b F_l^-}{l+1} \, \Big[ l (h_2-h_3) + 2h_2 \Big] \,,
\nonumber\\
G_M &=& - \frac{2b F_l^-}{l+1} \, (h_2-h_3) \,,
\\
G_E &=& 2b F_l^- \, h_1 \nonumber
\en
for normal parity transitions.
Here
\eq
F_l^\pm = \frac{M_2}{|{\bf p}|^l} \, \sqrt{c_{l+1} \,
\frac{M_2 E}{6 \pi \alpha M_1} \, \frac{Q_\pm}{M_\pm^2}} \,.
\en
   
The form factors $G_i^{(n)}(Q^2)$ (here $n$ is the radial quantum number),
defining the abnormal parity transitions, are given by 
\eq
G_1^{(n)}(Q^2) &=& \sum\limits_{\tau\tau^*} \,
\biggl[  g_{\tau\tau^*}^{(2A)} L_{1-\tau\tau^*}^{(n)}(Q^2)
       - g_{\tau\tau^*}^{(1D)} L_{1+\tau\tau^*}^{(n)}(Q^2)
       + g_{\tau\tau^*}^{(3A)} L_{2+\tau\tau^*}^{(n)}(Q^2)
       - g_{\tau\tau^*}^{(4D)} L_{2-\tau\tau^*}^{(n)}(Q^2)
\biggr]
\,, \nonumber\\
G_2^{(n)}(Q^2) &=&  \frac{1}{\kappa} \, \sum\limits_{\tau\tau^*} \,
\biggl[- g_{\tau\tau^*}^{(1B)} L_{2-\tau\tau^*}^{(n)}(Q^2)
       + g_{\tau\tau^*}^{(4B)} L_{1+\tau\tau^*}^{(n)}(Q^2)
       + g_{\tau\tau^*}^{(2E)} L_{2+\tau\tau^*}^{(n)}(Q^2)
       - g_{\tau\tau^*}^{(3E)} L_{1-\tau\tau^*}^{(n)}(Q^2)
\biggr] \,, \nonumber\\
G_3^{(n)}(Q^2) &=&  \frac{1}{\kappa} \, \sum\limits_{\tau\tau^*} \,
\biggl[  g_{\tau\tau^*}^{(2A)} M_{2+\tau\tau^*}^{(n)}(Q^2)
       - g_{\tau\tau^*}^{(1D)} M_{2-\tau\tau^*}^{(n)}(Q^2)
       - g_{\tau\tau^*}^{(3A)} M_{1-\tau\tau^*}^{(n)}(Q^2)
       + g_{\tau\tau^*}^{(4D)} M_{1+\tau\tau^*}^{(n)}(Q^2)
\nonumber\\
&+&      g_{\tau\tau^*}^{(1B)} K_{2-\tau\tau^*}^{(n)}(Q^2)
       - g_{\tau\tau^*}^{(4B)} K_{1+\tau\tau^*}^{(n)}(Q^2)
       - g_{\tau\tau^*}^{(2E)} K_{2+\tau\tau^*}^{(n)}(Q^2)
       + g_{\tau\tau^*}^{(3E)} K_{1-\tau\tau^*}^{(n)}(Q^2)
\nonumber\\
&+&      g_{\tau\tau^*}^{(1C)} R_{2-\tau\tau^*}^{(n)}(Q^2)
       - g_{\tau\tau^*}^{(4C)} R_{1+\tau\tau^*}^{(n)}(Q^2)
       - g_{\tau\tau^*}^{(2F)} R_{2+\tau\tau^*}^{(n)}(Q^2)
       + g_{\tau\tau^*}^{(3F)} R_{1-\tau\tau^*}^{(n)}(Q^2)
\biggr]
\,.
\en
\eq 
G_1^{(n)}(Q^2) &=& \sum\limits_{\tau\tau^*} \, 
\biggl[  g_{\tau\tau^*}^{(1A)} L_{1+\tau\tau^*}^{(n)}(Q^2) 
       - g_{\tau\tau^*}^{(2D)} L_{1-\tau\tau^*}^{(n)}(Q^2) 
       + g_{\tau\tau^*}^{(4A)} L_{2-\tau\tau^*}^{(n)}(Q^2)
       - g_{\tau\tau^*}^{(3D)} L_{2+\tau\tau^*}^{(n)}(Q^2)
\biggr] \,, \nonumber\\ 
G_2^{(n)}(Q^2) &=& \frac{1}{\kappa} \, \sum\limits_{\tau\tau^*} \,
\biggl[- g_{\tau\tau^*}^{(2B)} L_{2+\tau\tau^*}^{(n)}(Q^2) 
       + g_{\tau\tau^*}^{(3B)} L_{1-\tau\tau^*}^{(n)}(Q^2) 
       + g_{\tau\tau^*}^{(1E)} L_{2-\tau\tau^*}^{(n)}(Q^2)
       - g_{\tau\tau^*}^{(4E)} L_{1+\tau\tau^*}^{(n)}(Q^2)
\biggr] \,, \nonumber\\ 
G_3^{(n)}(Q^2) &=&  \frac{1}{\kappa} \, \sum\limits_{\tau\tau^*} \,
\biggl[  g_{\tau\tau^*}^{(1A)} M_{2-\tau\tau^*}^{(n)}(Q^2) 
       - g_{\tau\tau^*}^{(2D)} M_{2+\tau\tau^*}^{(n)}(Q^2) 
       - g_{\tau\tau^*}^{(4A)} M_{1+\tau\tau^*}^{(n)}(Q^2)
       + g_{\tau\tau^*}^{(3D)} M_{1-\tau\tau^*}^{(n)}(Q^2)
\nonumber\\
&+&      g_{\tau\tau^*}^{(2B)} K_{2+\tau\tau^*}^{(n)}(Q^2) 
       - g_{\tau\tau^*}^{(3B)} K_{1-\tau\tau^*}^{(n)}(Q^2) 
       - g_{\tau\tau^*}^{(1E)} K_{2-\tau\tau^*}^{(n)}(Q^2)
       + g_{\tau\tau^*}^{(4E)} K_{1+\tau\tau^*}^{(n)}(Q^2)             
\nonumber\\
&+&      g_{\tau\tau^*}^{(2C)} R_{2+\tau\tau^*}^{(n)}(Q^2) 
       - g_{\tau\tau^*}^{(3C)} R_{1-\tau\tau^*}^{(n)}(Q^2) 
       - g_{\tau\tau^*}^{(1F)} R_{2-\tau\tau^*}^{(n)}(Q^2)
       + g_{\tau\tau^*}^{(4F)} R_{1+\tau\tau^*}^{(n)}(Q^2)             
\biggr]
\,, 
\en 
where $F_{i\pm\tau\tau^*}^{(n)}(Q^2)$ with $F=K,L,M,N,R$
are functions calculated in soft-wall model.
Functions $F_{i\pm\tau\tau^*}^{(n)}(Q^2)$ are written as
\eq
& &F_{1\pm\tau\tau^*}^{(n)}(Q^2) = \frac{1}{2} \,
\biggl[F^{(n)}(Q^2,\tau^*,\tau)   \pm  F^{(n)}(Q^2,\tau^*+1,\tau+1)\biggr]\,, 
\nonumber\\
& &F_{2\pm\tau\tau^*}^{(n)}(Q^2) = \frac{1}{2} \,
\biggl[F^{(n)}(Q^2,\tau^*,\tau+1) \pm  F^{(n)}(Q^2,\tau^*+1,\tau)\biggr]\,,
\en
and
\eq
R_{i\pm\tau\tau^*}^{(n)}(Q^2) = \frac{1}{\kappa} \biggl[
L_{i\pm\tau\tau^*}^{(n)}(Q^2) + N_{i\pm\tau\tau^*}^{(n)}(Q^2) \biggr]\,.
\en
For $n=0$ one gets 
\eq
K^{(0)}(Q^2,\tau^*,\tau) &=& \frac{\Big(\tau^* - \frac{3}{2}\Big)
\, K(a,\tau^*,\tau)
-  K(a,\tau^* +2,\tau)}{2 \sqrt{\Gamma(\tau^* - 1) \Gamma(\tau-1)}} 
\,, \nonumber\\
L^{(0)}(Q^2,\tau^*,\tau) &=& \frac{ L(a,\tau^*,\tau)}{\sqrt{\Gamma(\tau^*-1) 
\Gamma(\tau-1)}} 
\,, \\
M^{(0)}(Q^2,\tau^*,\tau) &=&
\frac{M(a,\tau^*,\tau)}{2 \sqrt{\Gamma(\tau^*-1) \Gamma(\tau-1)}} 
\,, \nonumber\\
N^{(0)}(Q^2,\tau^*,\tau) &=&
\frac{K(a,\tau^*,\tau) -  2 K(a,\tau^*+2,\tau)}
{2 \sqrt{\Gamma(\tau^*-1) \Gamma(\tau-1)}}  \,. \nonumber 
\en
Here
\eq
K(a,\tau^*,\tau)&=& \Gamma\Big(\frac{\tau^* + \tau}{2}\Big) \,
                    B\Big(a+1,\frac{ \tau^* + \tau}{2}\Big) \,,
\nonumber\\
L(a,\tau^*,\tau)&=& \Gamma\Big(\frac{\tau^* + \tau + 2}{2}\Big) \,
                    B\Big(a+1,\frac{ \tau^* + \tau    }{2}\Big) \,,
\nonumber\\
M(a,\tau^*,\tau)&=& \Gamma\Big(\frac{\tau^* + \tau + 1}{2}\Big) \,
                    B\Big(a+1,\frac{ \tau^* + \tau + 1}{2}\Big) \,,
\nonumber\\
N(a,\tau^*,\tau)&=& K(a,\tau^*,\tau) - 2 K(a,\tau^*+2,\tau) \,.
\en

\begin{figure}[htb]
\begin{center}
\epsfig{figure=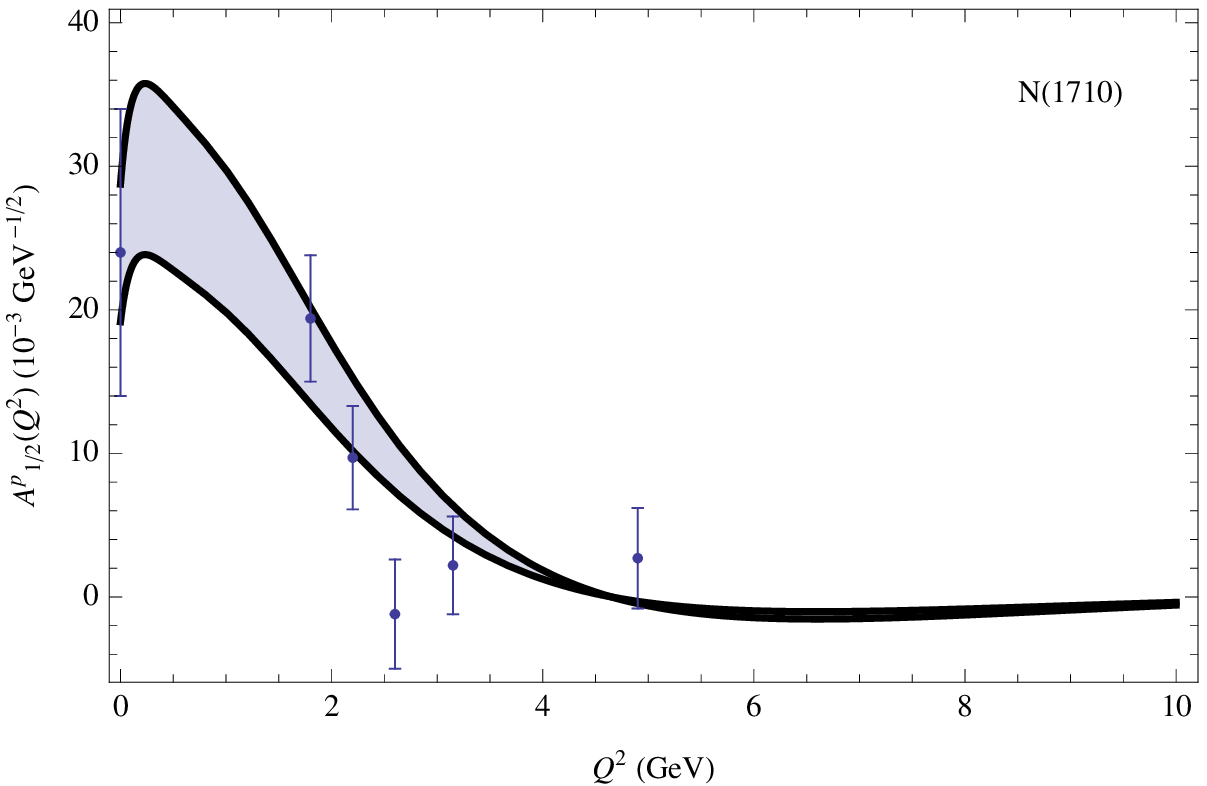,scale=.7}
\epsfig{figure=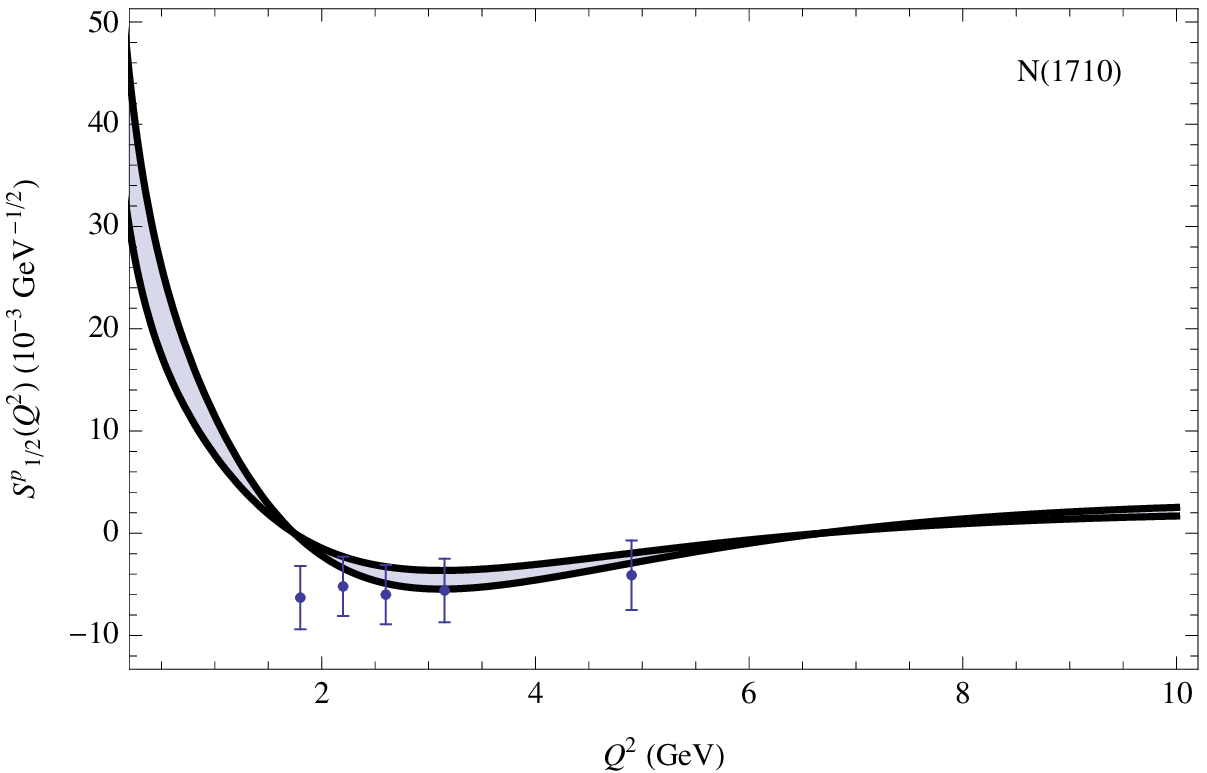,scale=.7}
\caption{Helicity amplitudes 
$A_{1/2}^p(Q^2)$ (left panel) and $S_{1/2}^p(Q^2)$ (right panel)  
for $N \gamma^* \to N(1710)$ transition 
up to $Q^2=$ 10~GeV$^2$. Our results are shown with 
a variation of the parameters of our approach (shaded band), and 
comparing with data taken from the CLAS Collaboration~\cite{Park:2014yea}. 
Here and in the following superscript ($p$) in the notation of the 
helicity amplitudes means the proton channel. 
\label{fig1}}

\epsfig{figure=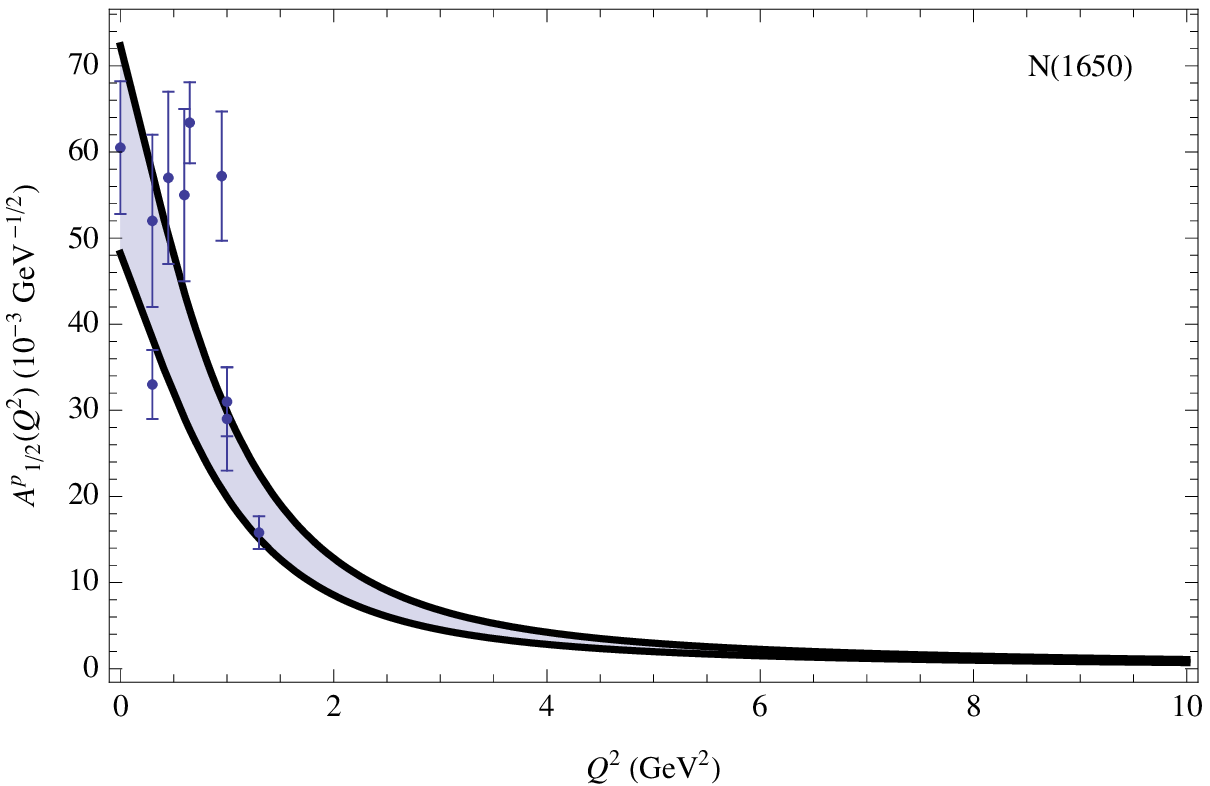,scale=.7}
\epsfig{figure=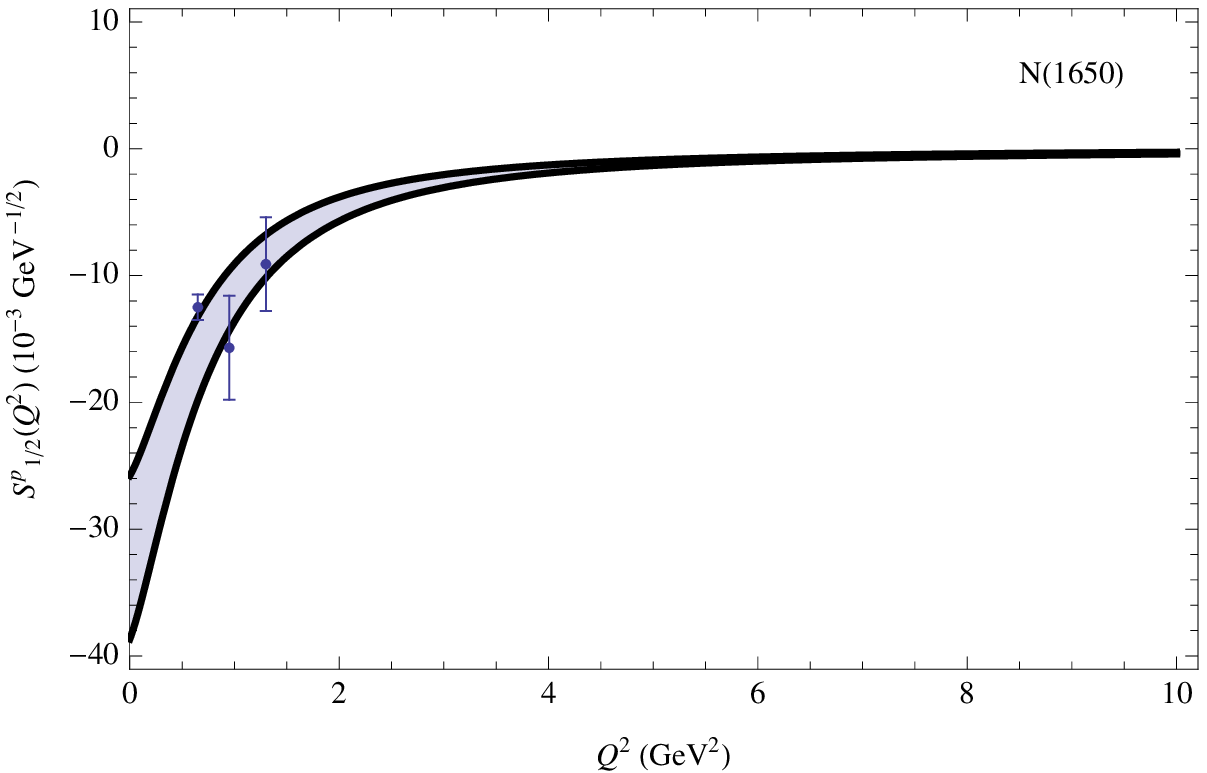,scale=.7}
\caption{Helicity amplitudes 
$A_{1/2}^p(Q^2)$ (left panel) and $S_{1/2}^p(Q^2)$ (right panel)  
for $N \gamma^* \to N(1650)$ transition 
up to $Q^2=$ 10~GeV$^2$. Our results are shown with a 
variation of the parameters of our approach (shaded band), 
comparing with data taken from the CLAS 
Collaboration~\cite{Mokeev:2013kka,Golovatch:2018hjk} and 
compilation of the world analyses of the $N\pi$ electroproduction 
data~\cite{Burkert:2002zz}. 
\label{fig2}}

\epsfig{figure=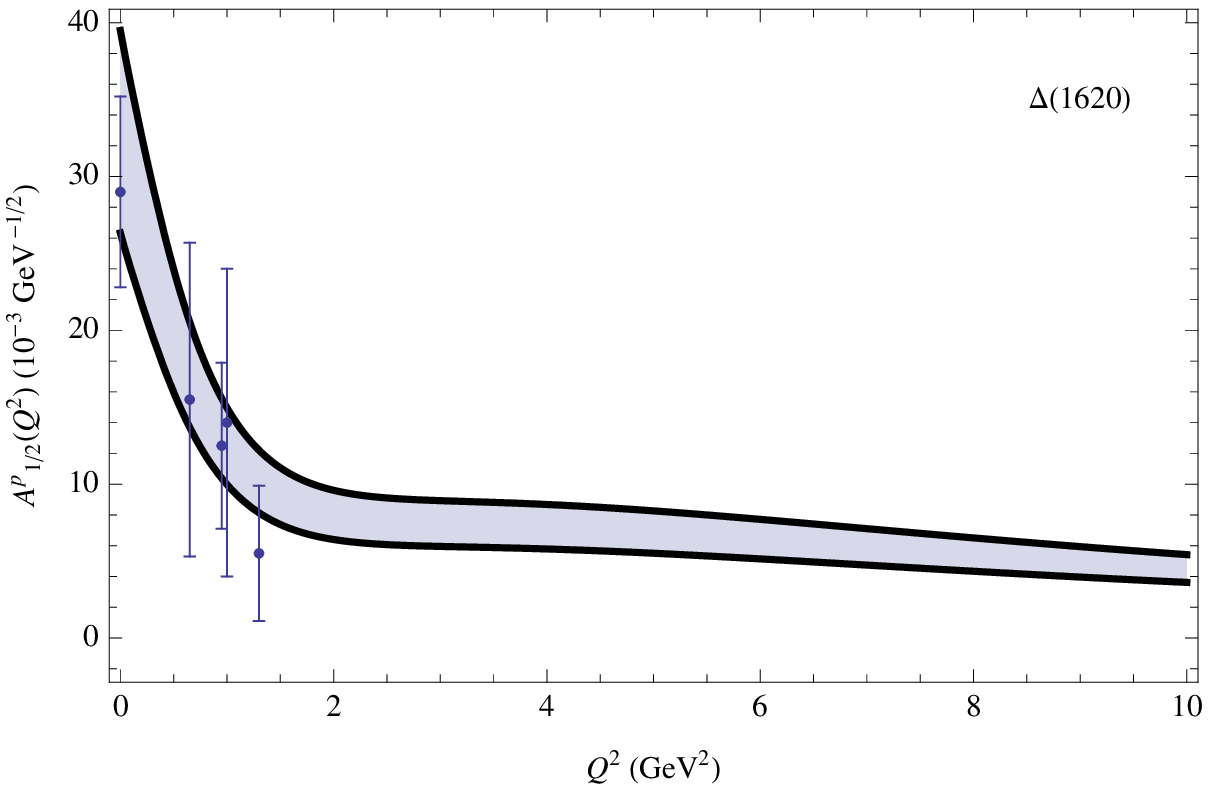,scale=.7}
\epsfig{figure=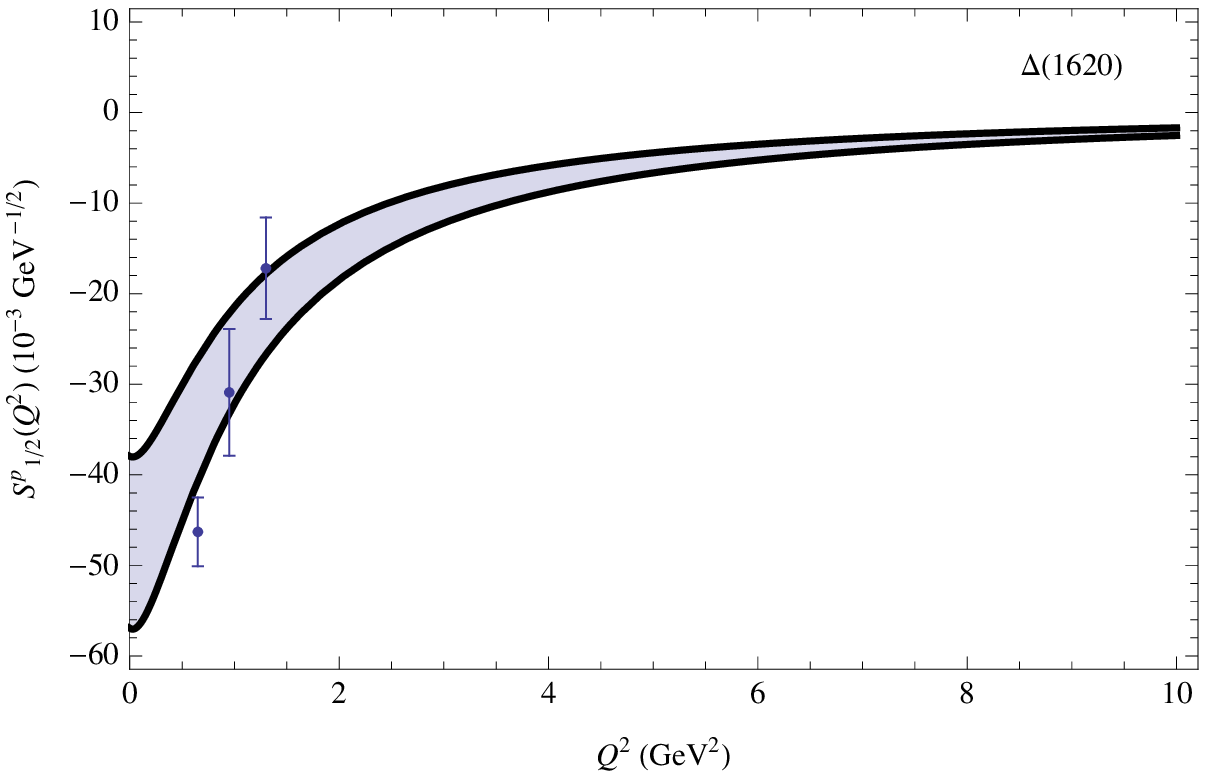,scale=.7}
\caption{Helicity amplitudes 
$A_{1/2}^p(Q^2)$ (left panel) and $S_{1/2}^p(Q^2)$ (right panel)  
for $N \gamma^* \to \Delta(1620)$ transition 
up to $Q^2=$ 10~GeV$^2$. Our results (shaded band) are compared 
with data taken from the CLAS 
Collaboration~\cite{Mokeev:2015lda,Golovatch:2018hjk},  
a compilation of the world analyses of $N\pi$ electroproduction 
data~\cite{Burkert:2002zz} and Particle Data Group (PDG)~\cite{PDG20}. 
\label{fig3}}
\end{center}
\end{figure}

\clearpage 

\begin{figure}[htb]
\begin{center}
\vspace*{-.5cm}
\epsfig{figure=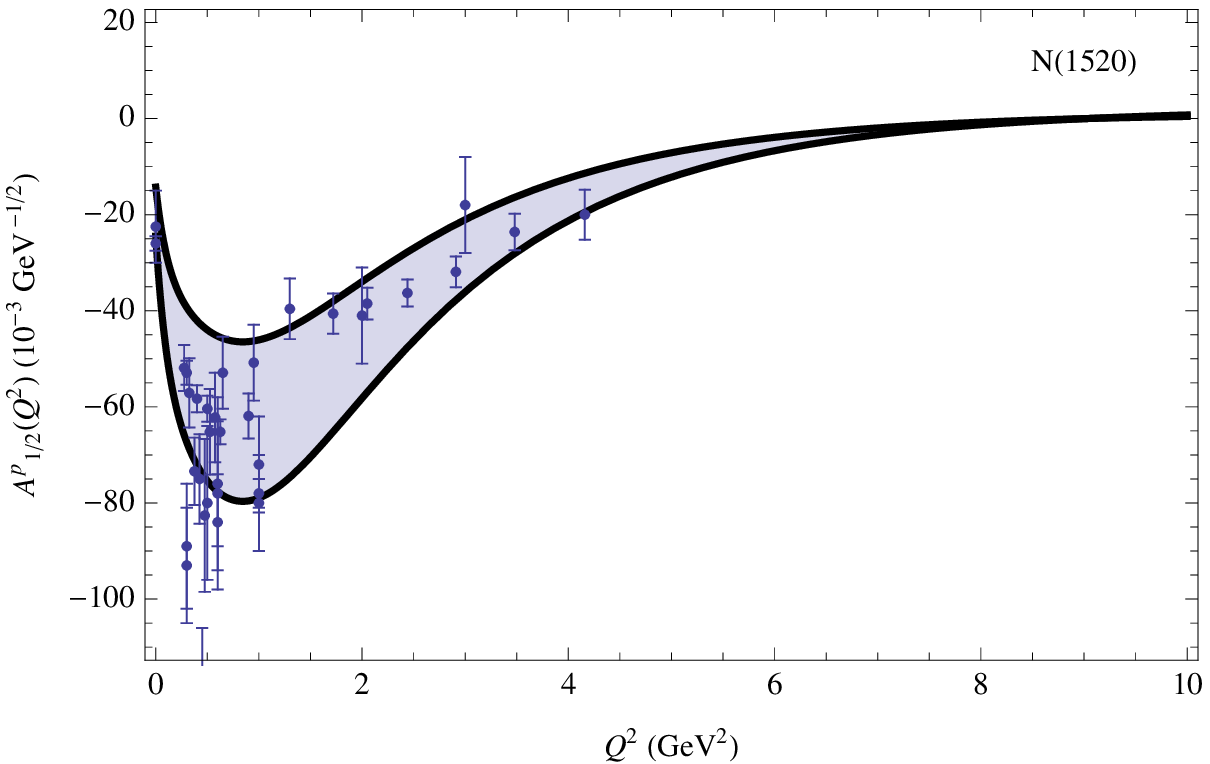,scale=.6}
\epsfig{figure=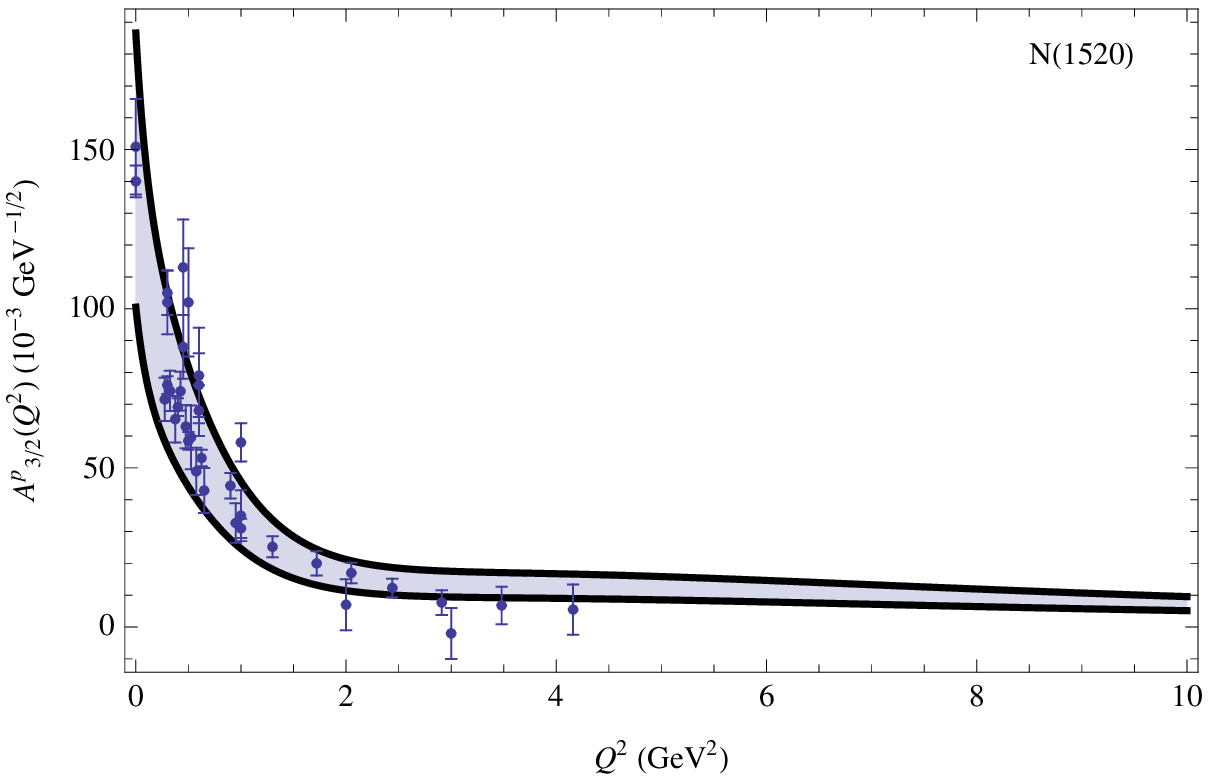,scale=.6}
\vspace*{.1cm}
\epsfig{figure=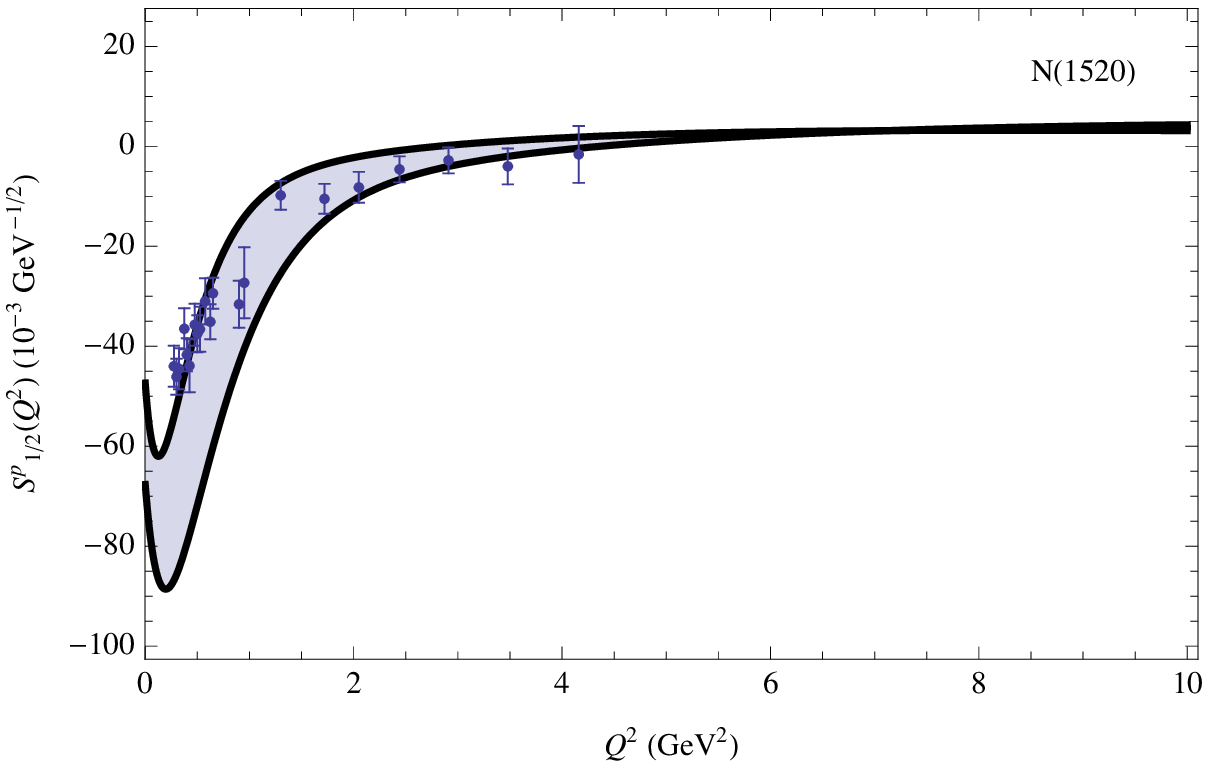,scale=.6}
\caption{Helicity amplitudes 
$A_{1/2}^p(Q^2)$ (left upper panel), 
$A_{3/2}^p(Q^2)$ (right upper panel), 
and 
$S_{1/2}^p(Q^2)$ (centered lower panel),   
for $N \gamma^* \to N(1520)$ transition 
up to $Q^2=$ 10~GeV$^2$. Our results (shaded band) are compared 
with data taken from the CLAS 
Collaboration~\cite{Dugger:2009pn,Aznauryan:2009mx,Mokeev:2012vsa,%
Mokeev:2015lda}, a compilation of data~\cite{Burkert:2002zz}, 
and PDG~\cite{PDG20}. 
\label{fig4}}

\epsfig{figure=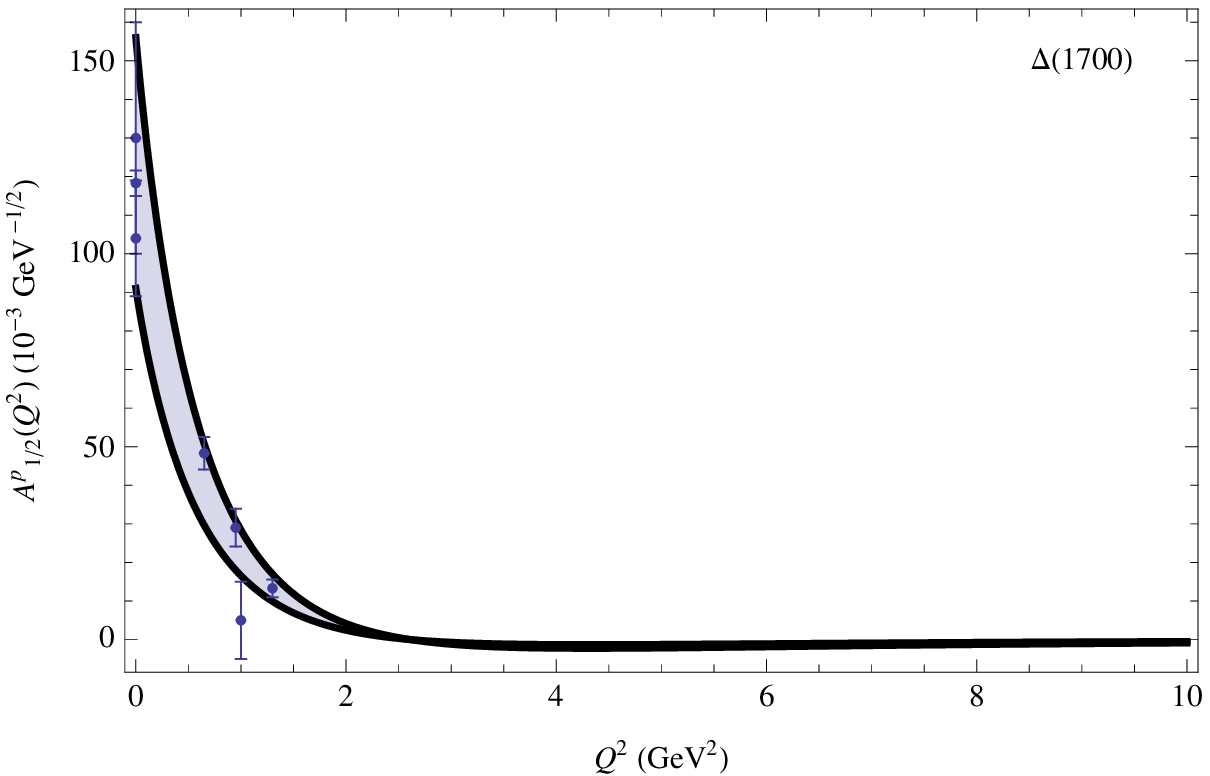,scale=.6}
\epsfig{figure=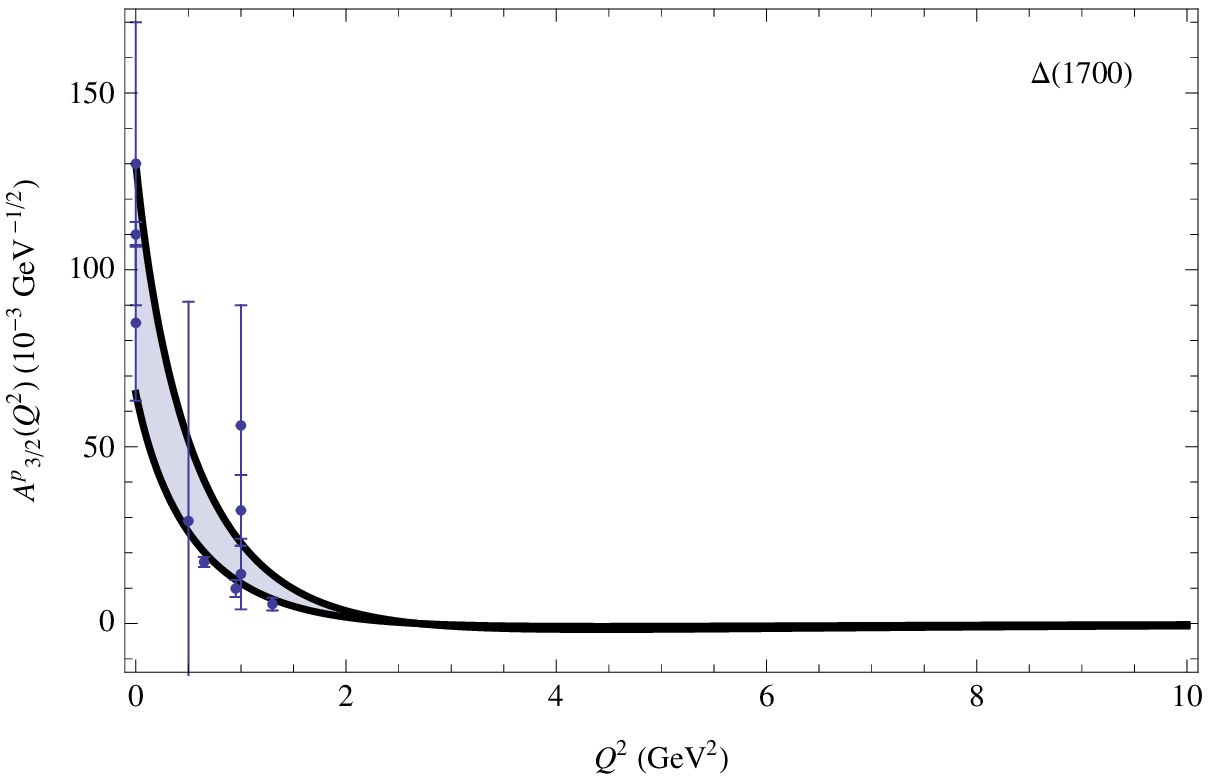,scale=.6}
\vspace*{.1cm}
\epsfig{figure=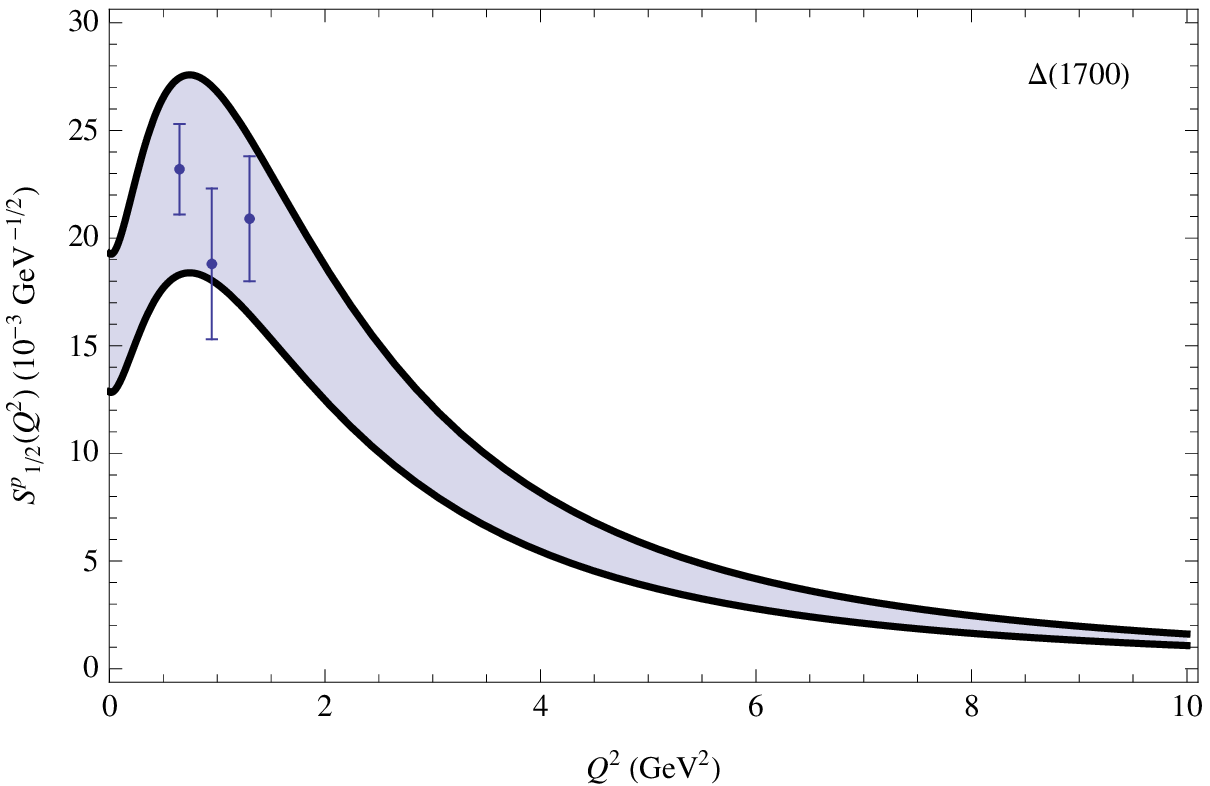,scale=.6}
\caption{Helicity amplitudes 
$A_{1/2}^p(Q^2)$ (left upper panel), 
$A_{3/2}^p(Q^2)$ (right upper panel), 
and 
$S_{1/2}^p(Q^2)$ (centered lower panel),   
for $N \gamma^* \to \Delta(1700)$ transition 
up to $Q^2=$ 10~GeV$^2$. Our results (shaded band) are compared 
with data taken from 
the CLAS Collaboration~\cite{Dugger:2009pn,Mokeev:2013kka}, 
a compilation of data~\cite{Burkert:2002zz}, and PDG~\cite{PDG20}. 
\label{fig5}}

\end{center}
\end{figure}

\clearpage 

\begin{figure}[htb]
\begin{center}
\vspace*{-.5cm}
\epsfig{figure=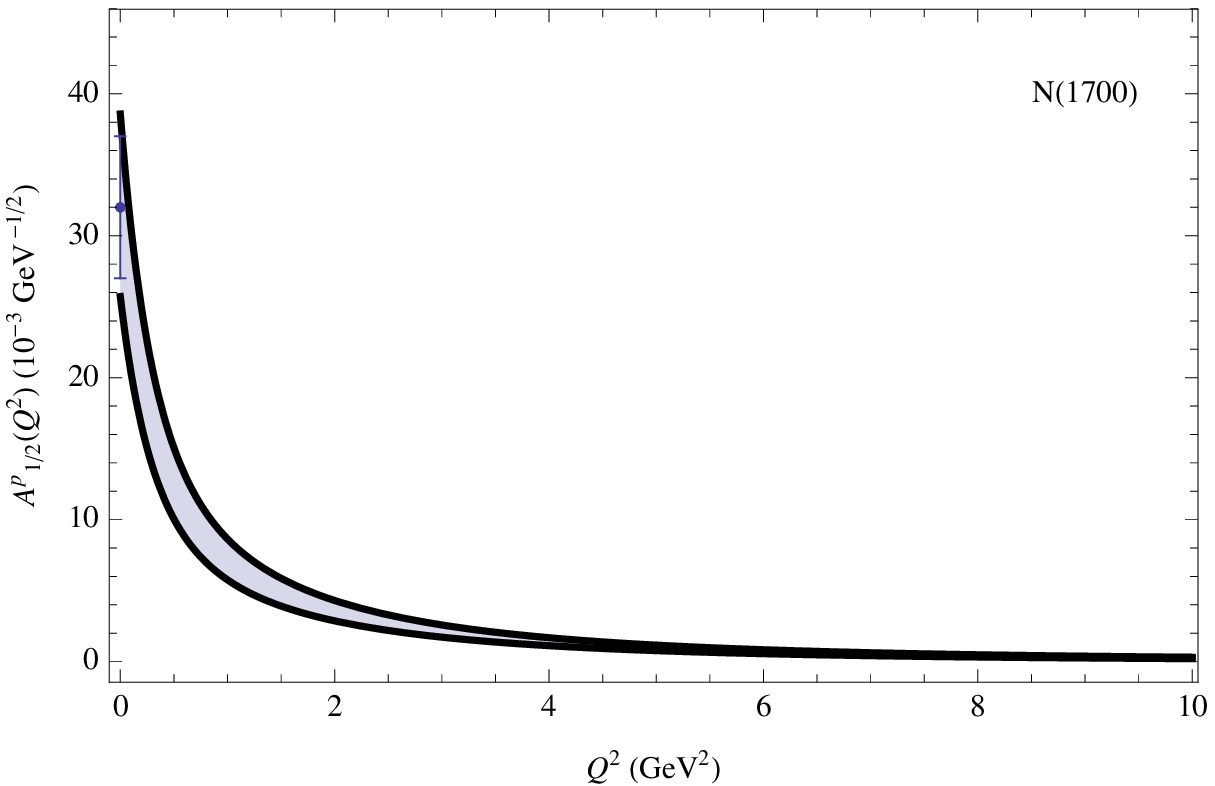,scale=.6}
\epsfig{figure=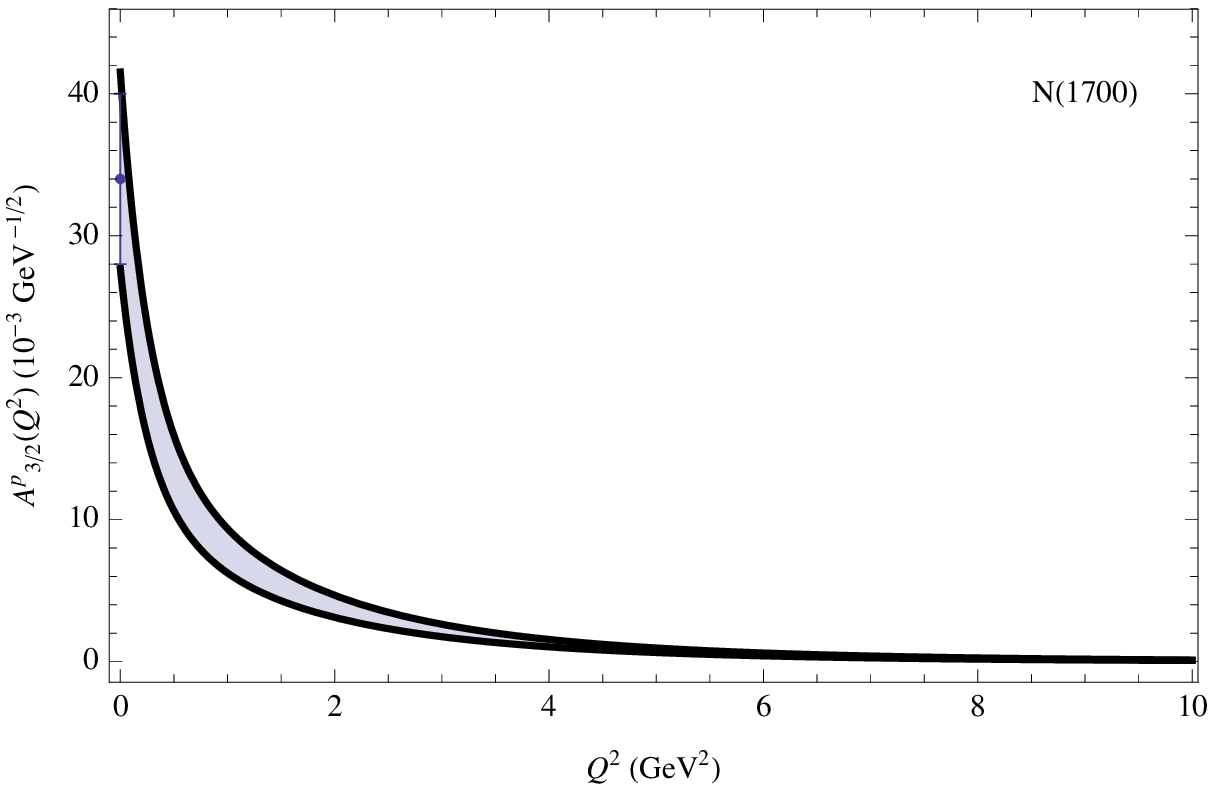,scale=.6}
\vspace*{.1cm}
\epsfig{figure=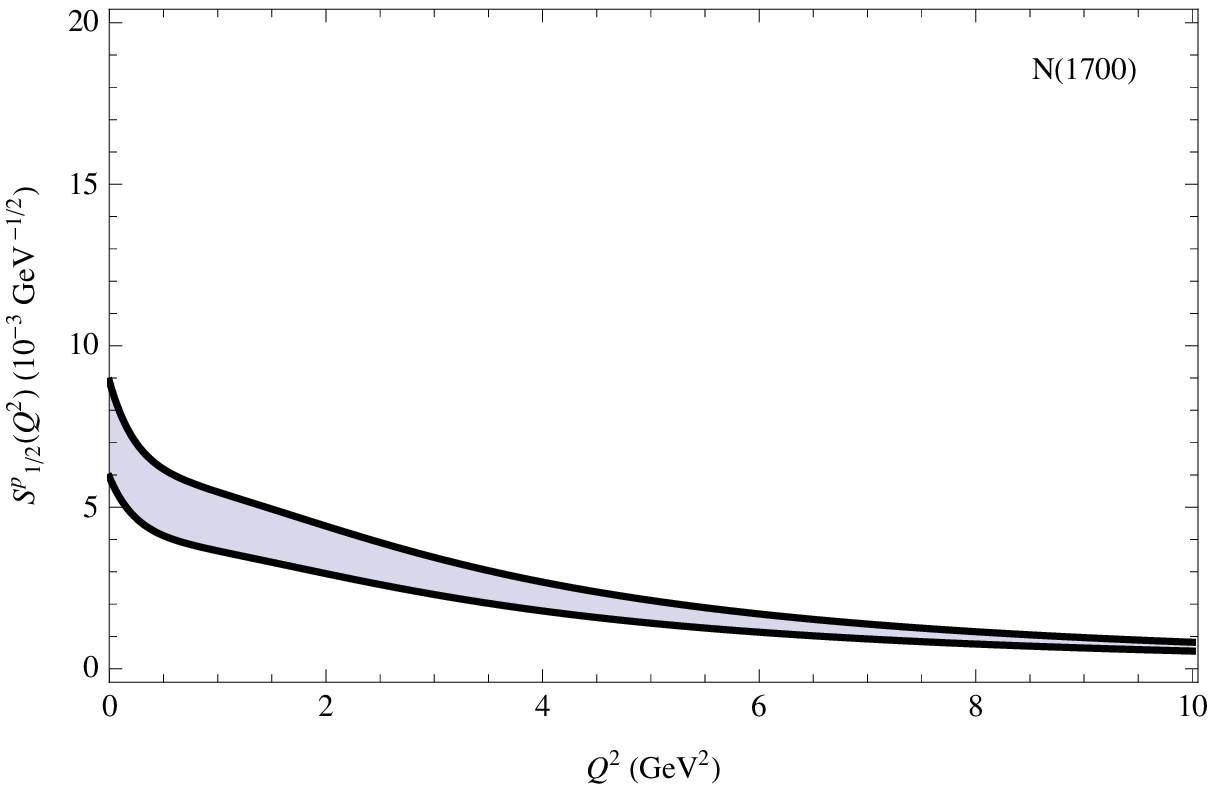,scale=.6}
\caption{Helicity amplitudes 
$A_{1/2}^p(Q^2)$ (left upper panel), 
$A_{3/2}^p(Q^2)$ (right upper panel), 
and 
$S_{1/2}^p(Q^2)$ (centered lower panel),   
for $N \gamma^* \to N(1700)$ transition 
up to $Q^2=$ 10~GeV$^2$. Our results (shaded band) are compared 
with PDG~\cite{PDG20}. 
\label{fig6}}

\epsfig{figure=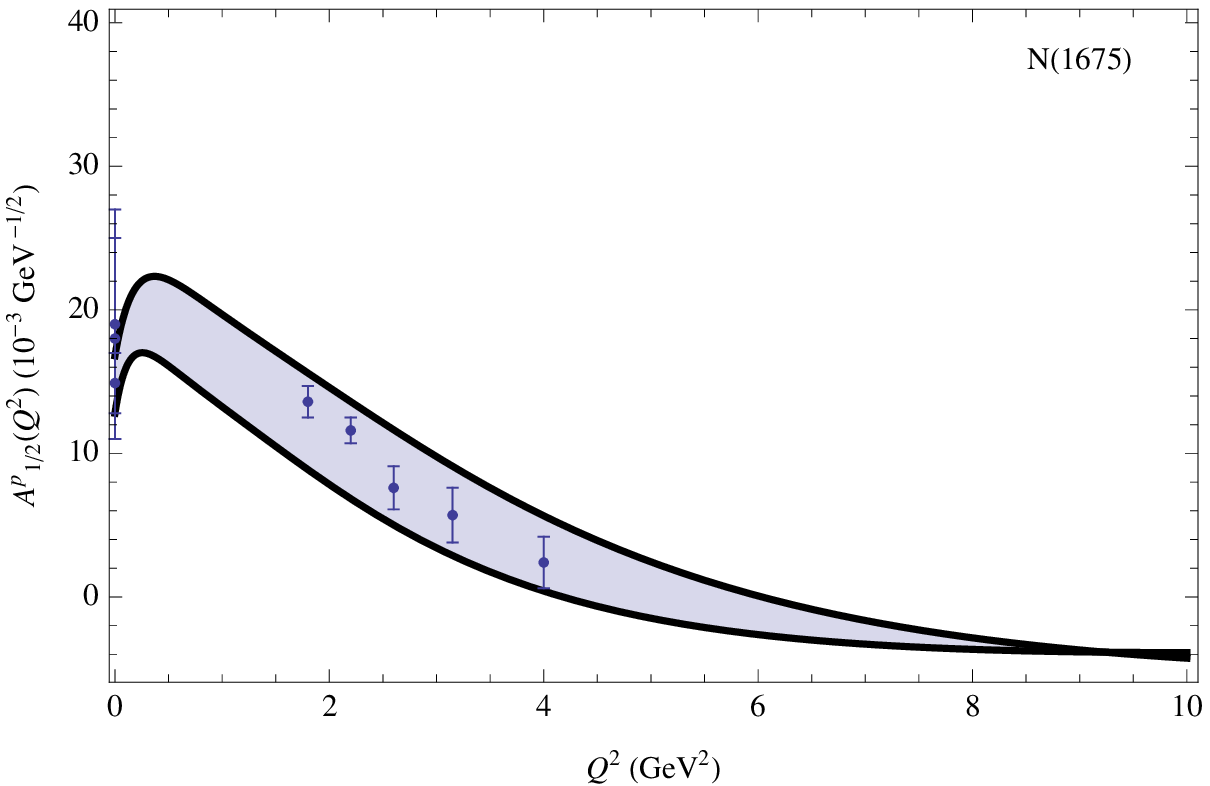,scale=.6}
\epsfig{figure=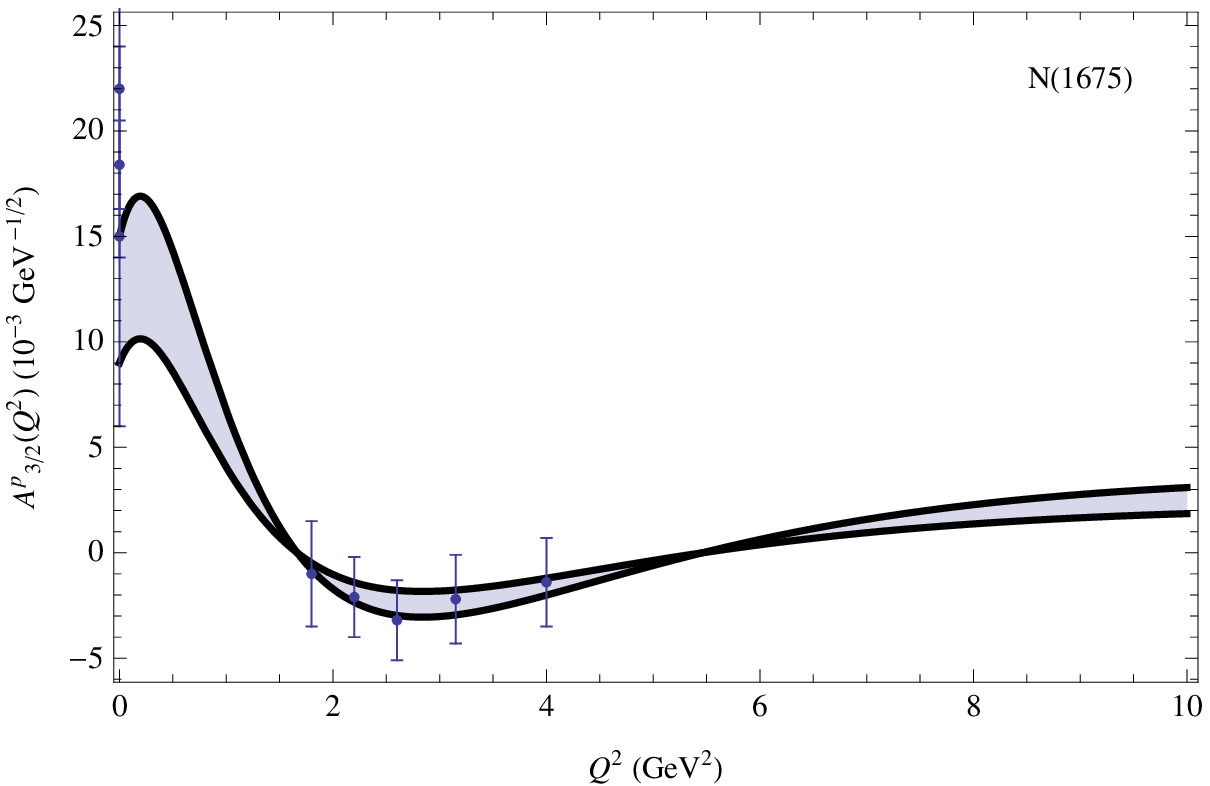,scale=.6}
\vspace*{.1cm}
\epsfig{figure=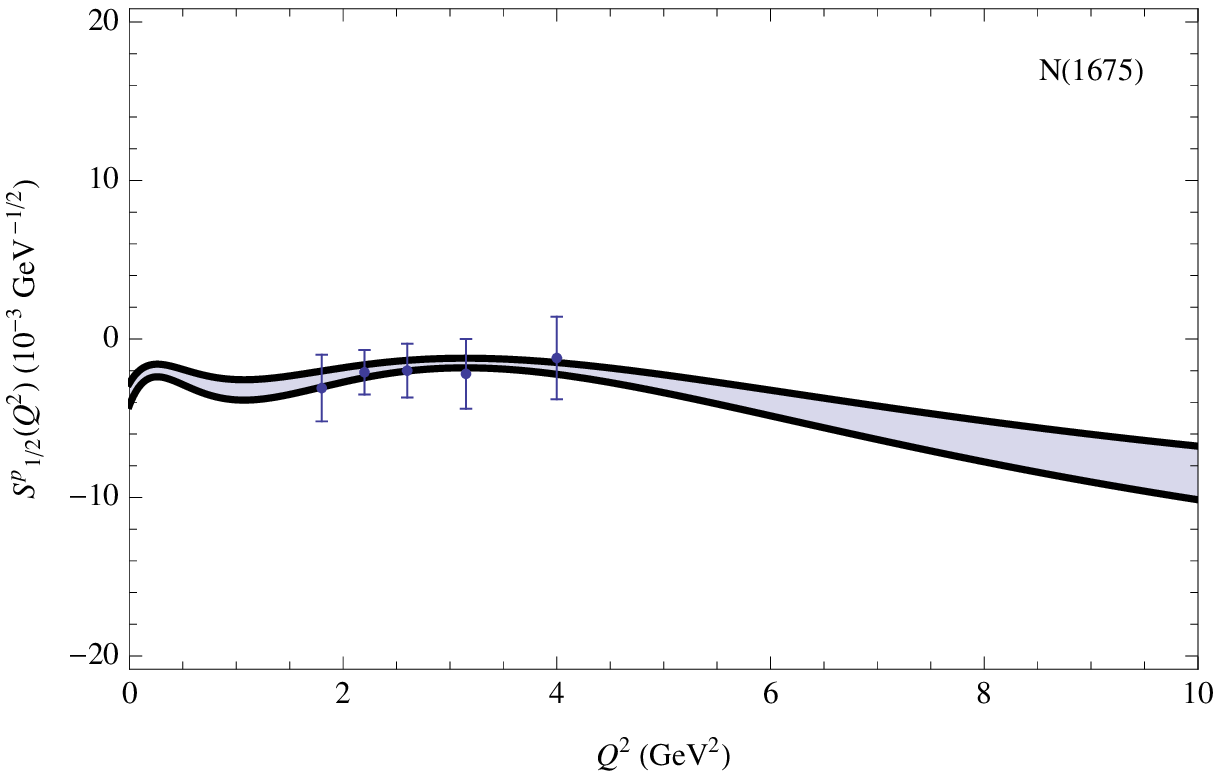,scale=.6}
\caption{Helicity amplitudes 
$A_{1/2}^p(Q^2)$ (left upper panel), 
$A_{3/2}^p(Q^2)$ (right upper panel), 
and 
$S_{1/2}^p(Q^2)$ (centered lower panel),   
for $N \gamma^* \to N(1675)$ transition 
up to $Q^2=$ 10~GeV$^2$. Our results (shaded band) 
are compared with data taken from the
CLAS Collaboration~\cite{Dugger:2009pn,Park:2014yea} 
and PDG~\cite{PDG20}. 
\label{fig7}}

\end{center}
\end{figure}

\clearpage 

\begin{figure}[htb]
\begin{center}
\vspace*{-.5cm}
\epsfig{figure=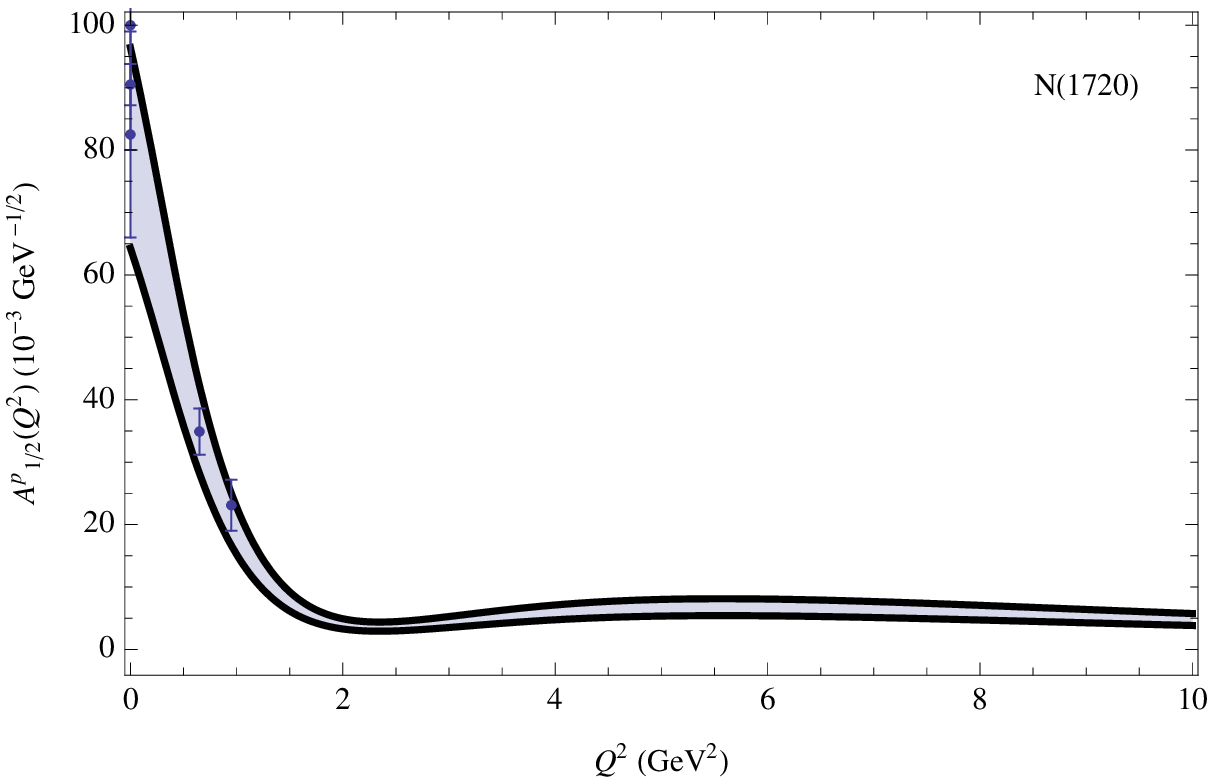,scale=.6}
\epsfig{figure=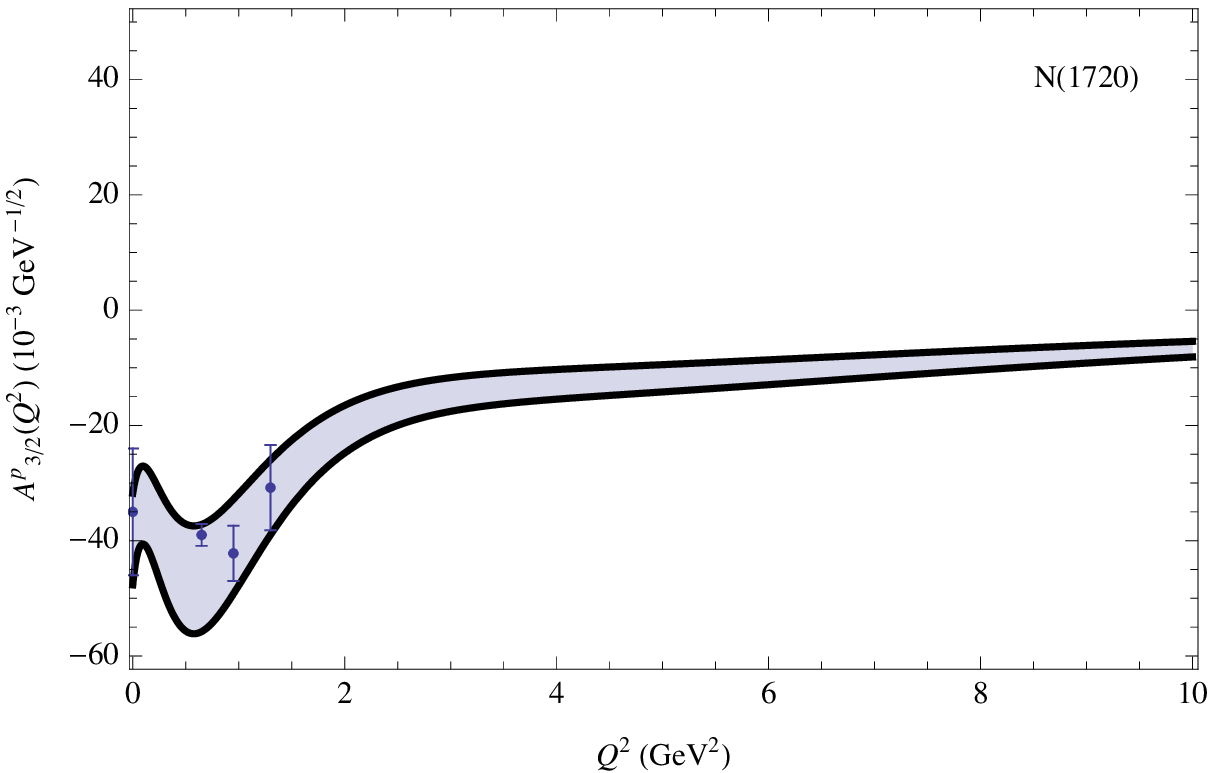,scale=.6}
\vspace*{.1cm}
\epsfig{figure=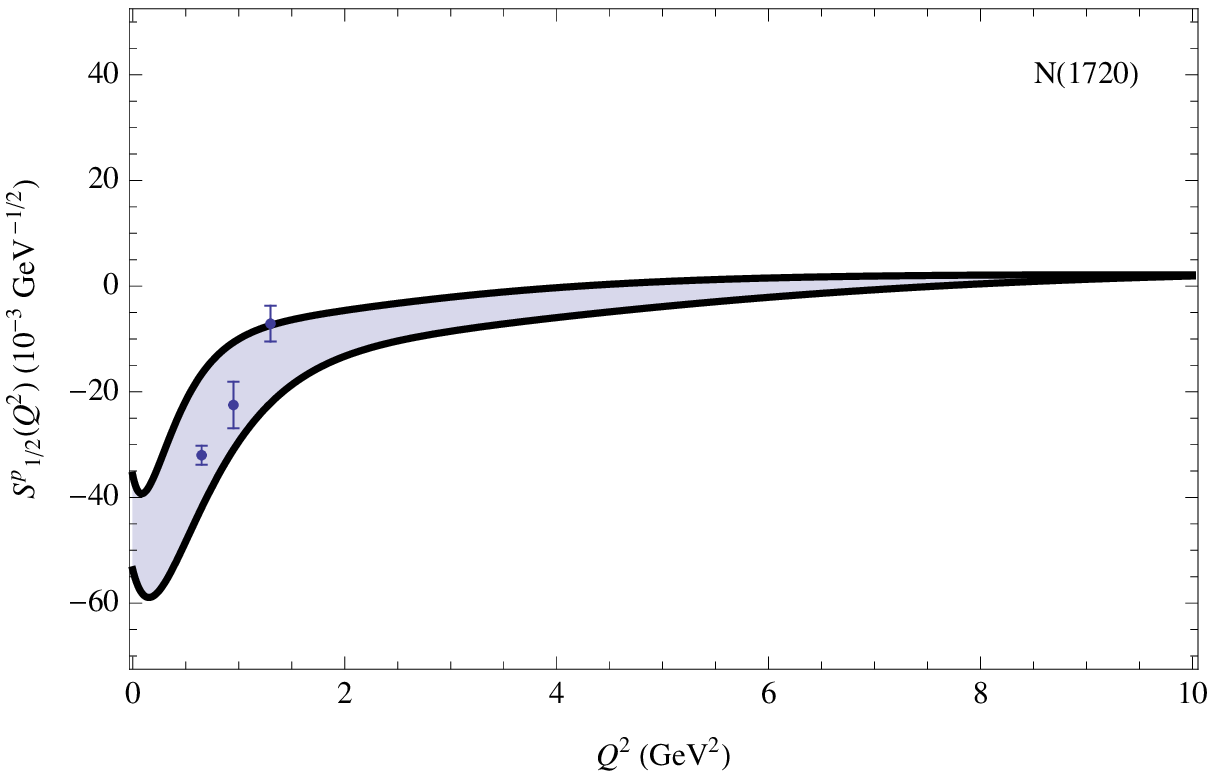,scale=.6}
\caption{Helicity amplitudes 
$A_{1/2}^p(Q^2)$ (left upper panel), 
$A_{3/2}^p(Q^2)$ (right upper panel), 
and 
$S_{1/2}^p(Q^2)$ (centered lower panel),   
for $N \gamma^* \to N(1720)$ transition 
up to $Q^2=$ 10~GeV$^2$. Our results (shaded band) 
are compared with data taken from the 
CLAS Collaboration~\cite{Mokeev:2020hhu} 
and PDG~\cite{PDG20}. 
\label{fig8}}

\epsfig{figure=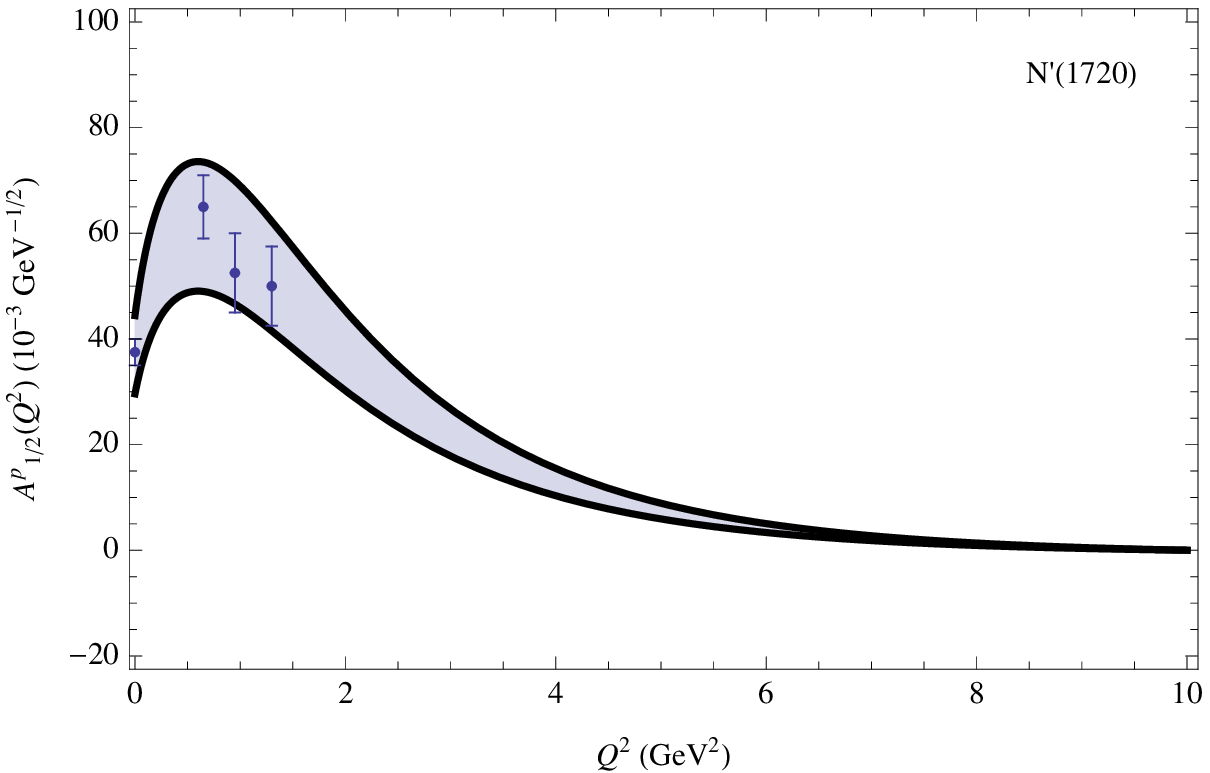,scale=.6}
\epsfig{figure=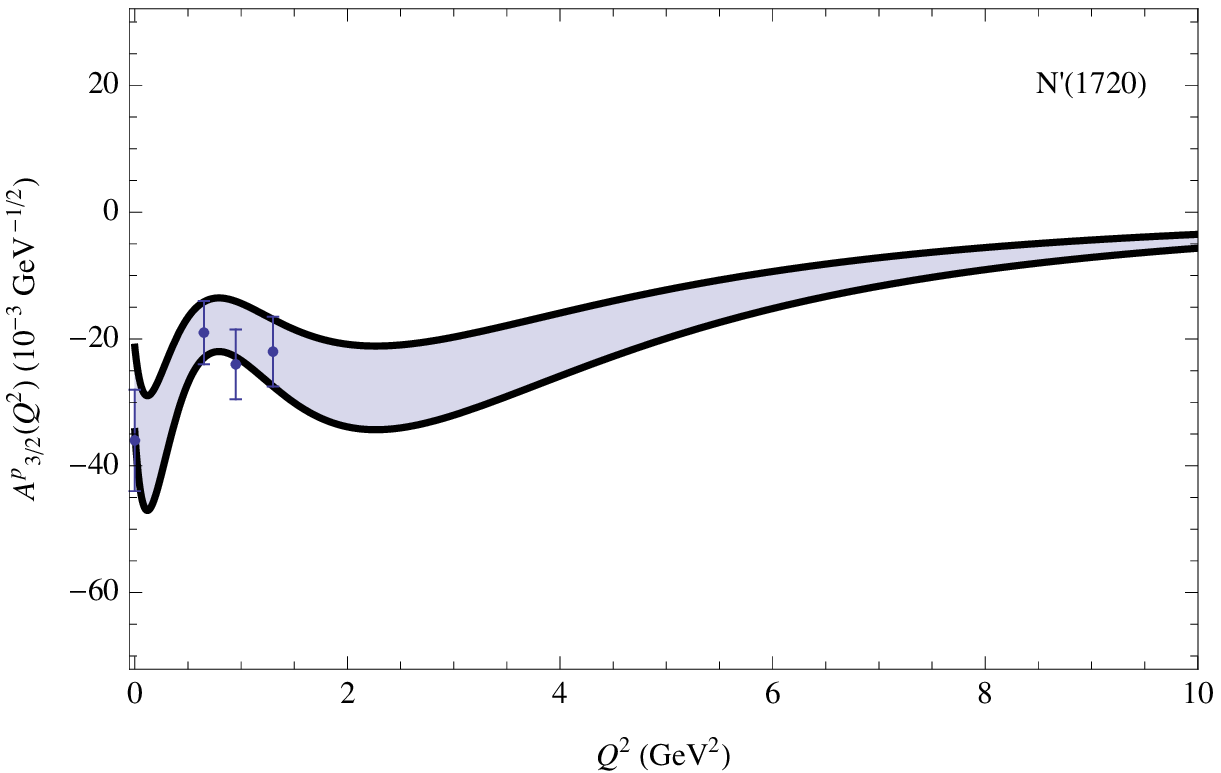,scale=.6}
\vspace*{.1cm}
\epsfig{figure=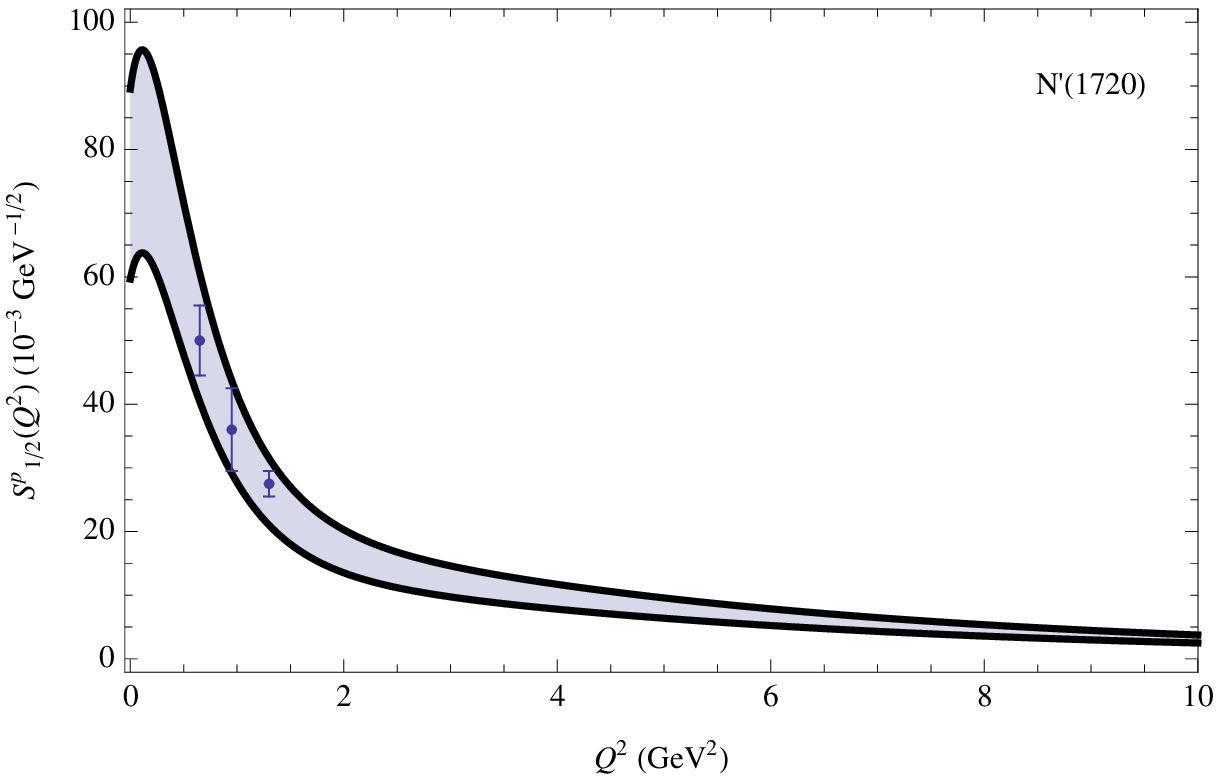,scale=.6}
\caption{Helicity amplitudes 
$A_{1/2}^p(Q^2)$ (left upper panel), 
$A_{3/2}^p(Q^2)$ (right upper panel), 
and 
$S_{1/2}^p(Q^2)$ (centered lower panel),   
for $N \gamma^* \to N'(1720)$ transition 
up to $Q^2=$ 10~GeV$^2$. Our results (shaded band) 
are compared with data taken from the
CLAS Collaboration~\cite{Mokeev:2020hhu} 
and PDG~\cite{PDG20}. 
\label{fig9}}

\end{center}
\end{figure}

\clearpage 

\begin{figure}[htb]
\begin{center}
\vspace*{-.5cm}
\epsfig{figure=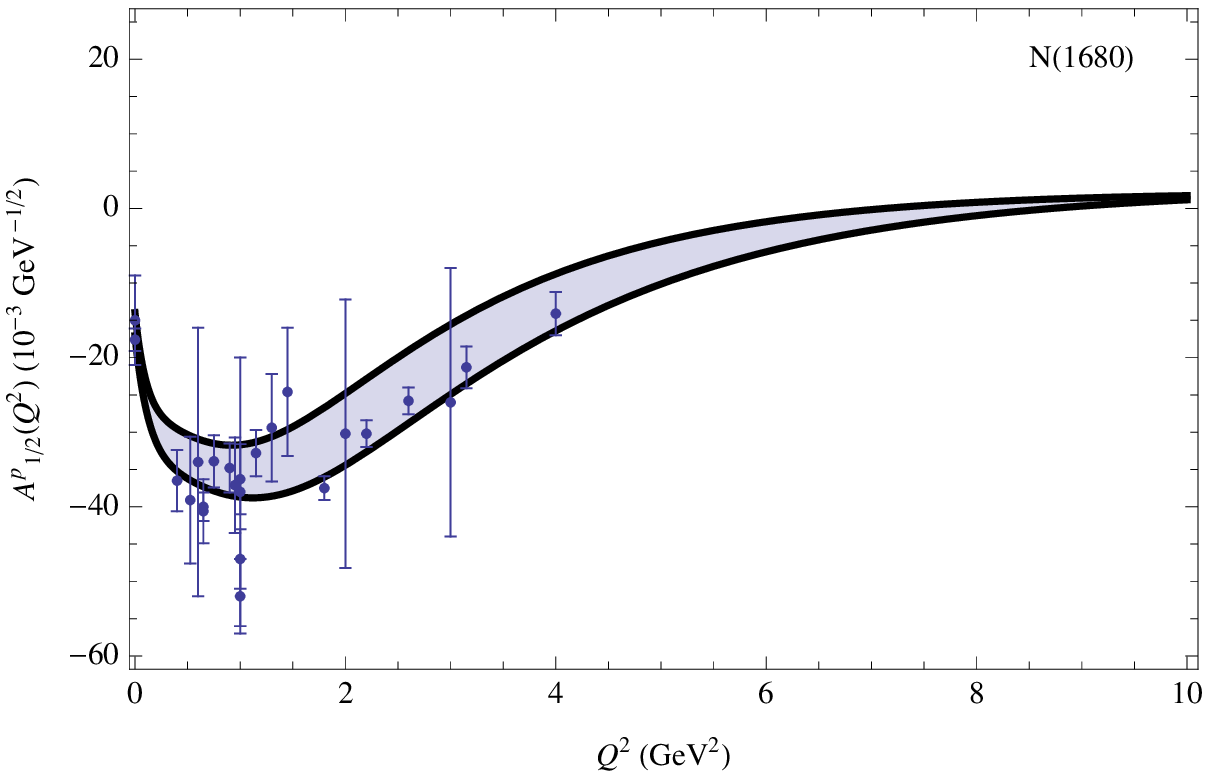,scale=.6}
\epsfig{figure=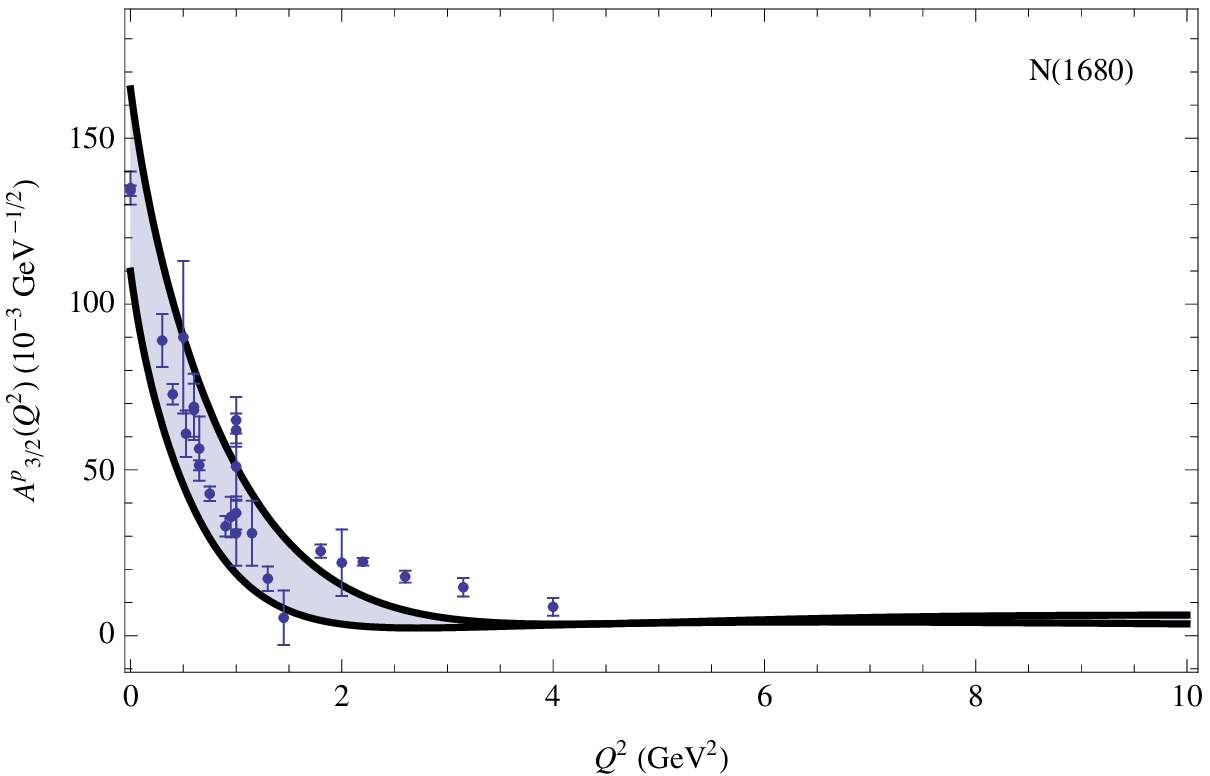,scale=.6}
\vspace*{.1cm}
\epsfig{figure=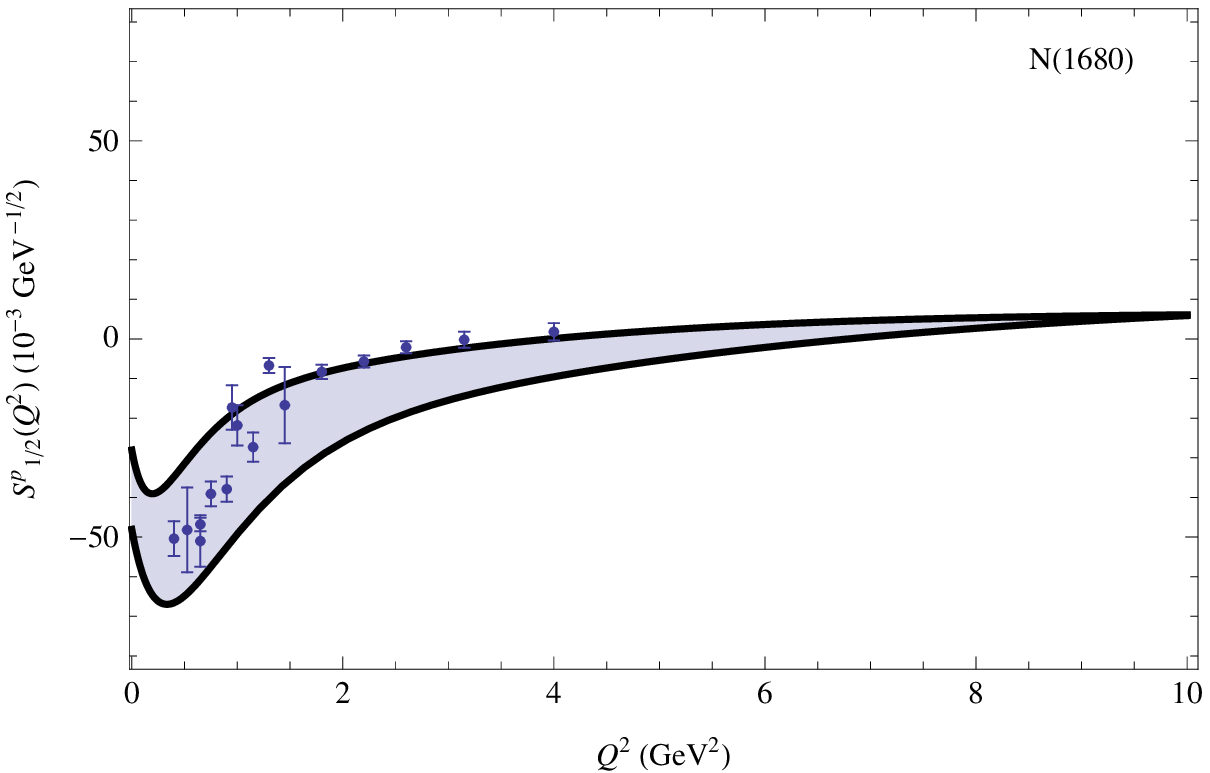,scale=.6}
\caption{Helicity amplitudes 
$A_{1/2}^p(Q^2)$ (left upper panel), 
$A_{3/2}^p(Q^2)$ (right upper panel), 
and 
$S_{1/2}^p(Q^2)$ (centered lower panel),   
for $N \gamma^* \to N(1680)$ transition 
up to $Q^2=$ 10~GeV$^2$. 
Our results (shaded band) are compared
with data taken from the
CLAS Collaboration~\cite{Dugger:2009pn,Park:2014yea,%
Mokeev:2013kka}, data analisys~\cite{Burkert:2002zz}  
and PDG~\cite{PDG20}. 
\label{fig10}}

\epsfig{figure=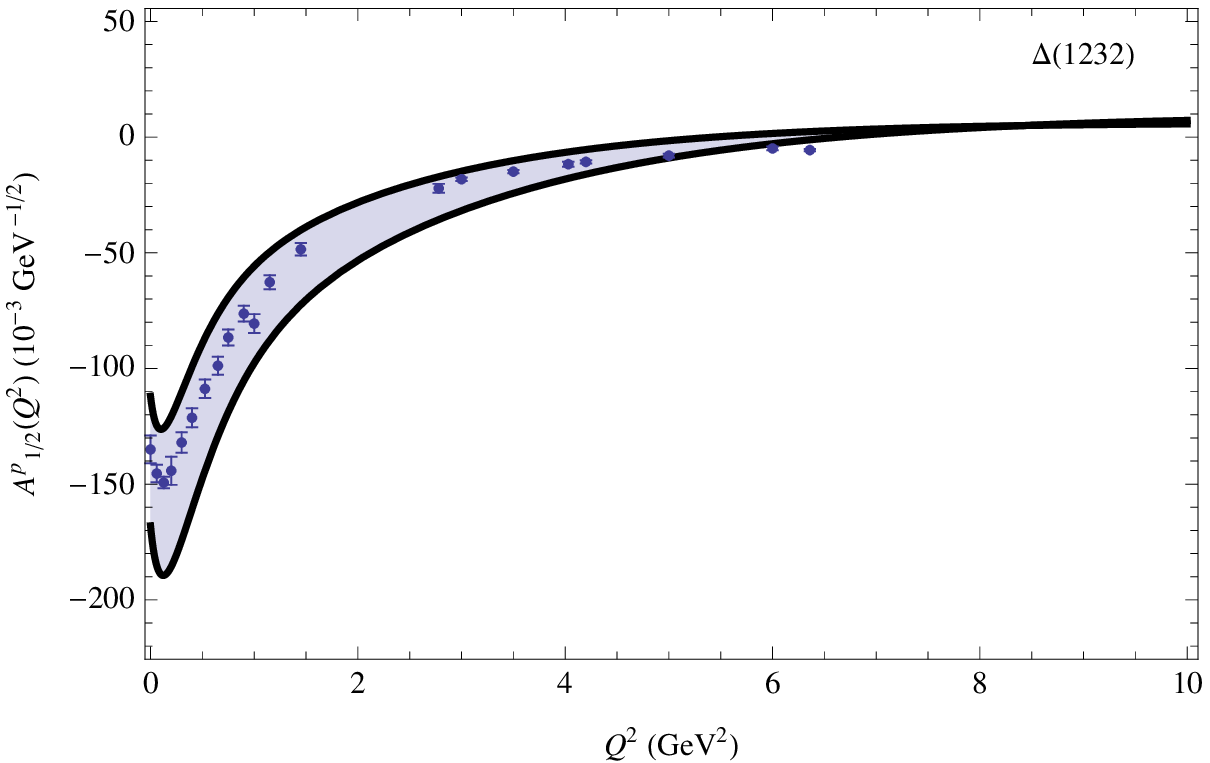,scale=.6}
\epsfig{figure=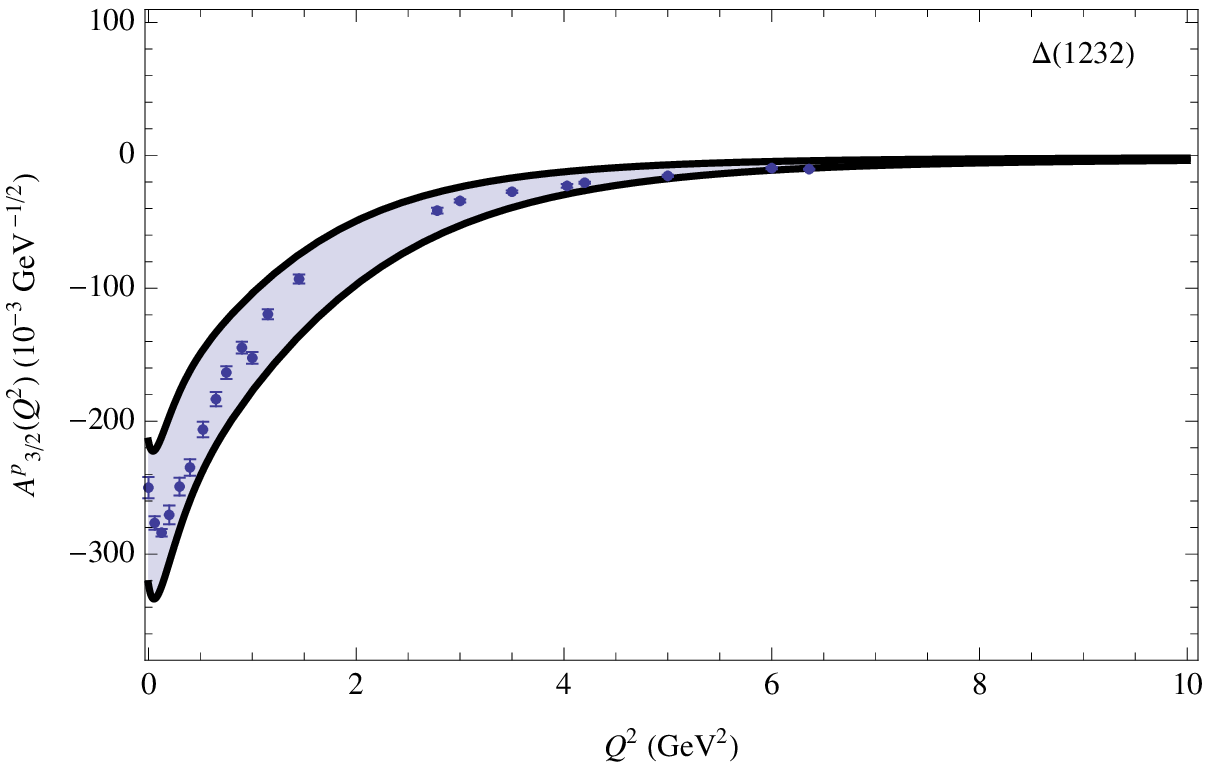,scale=.6}
\vspace*{.1cm}
\epsfig{figure=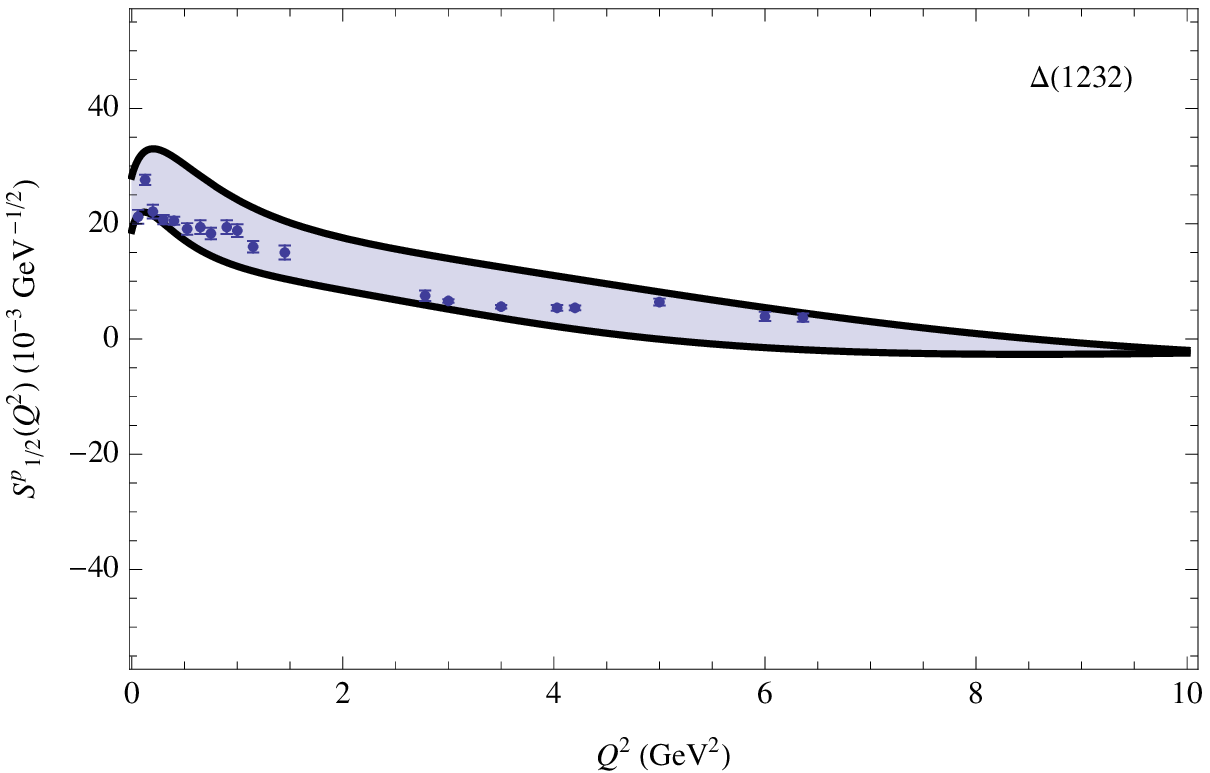,scale=.6}
\caption{Helicity amplitudes 
$A_{1/2}^p(Q^2)$ (left upper panel), 
$A_{3/2}^p(Q^2)$ (right upper panel), 
and 
$S_{1/2}^p(Q^2)$ (centered lower panel),   
for $N \gamma^* \to \Delta(1232)$ transition 
up to $Q^2=$ 10~GeV$^2$. 
Our results (shaded band) are compared
with data taken from the
CLAS Collaboration~\cite{Aznauryan:2009mx}   
and PDG~\cite{PDG20}. 
\label{fig11}}

\end{center}
\end{figure}

\clearpage 

\begin{figure}[htb]
\begin{center}
\vspace*{-.5cm}
\epsfig{figure=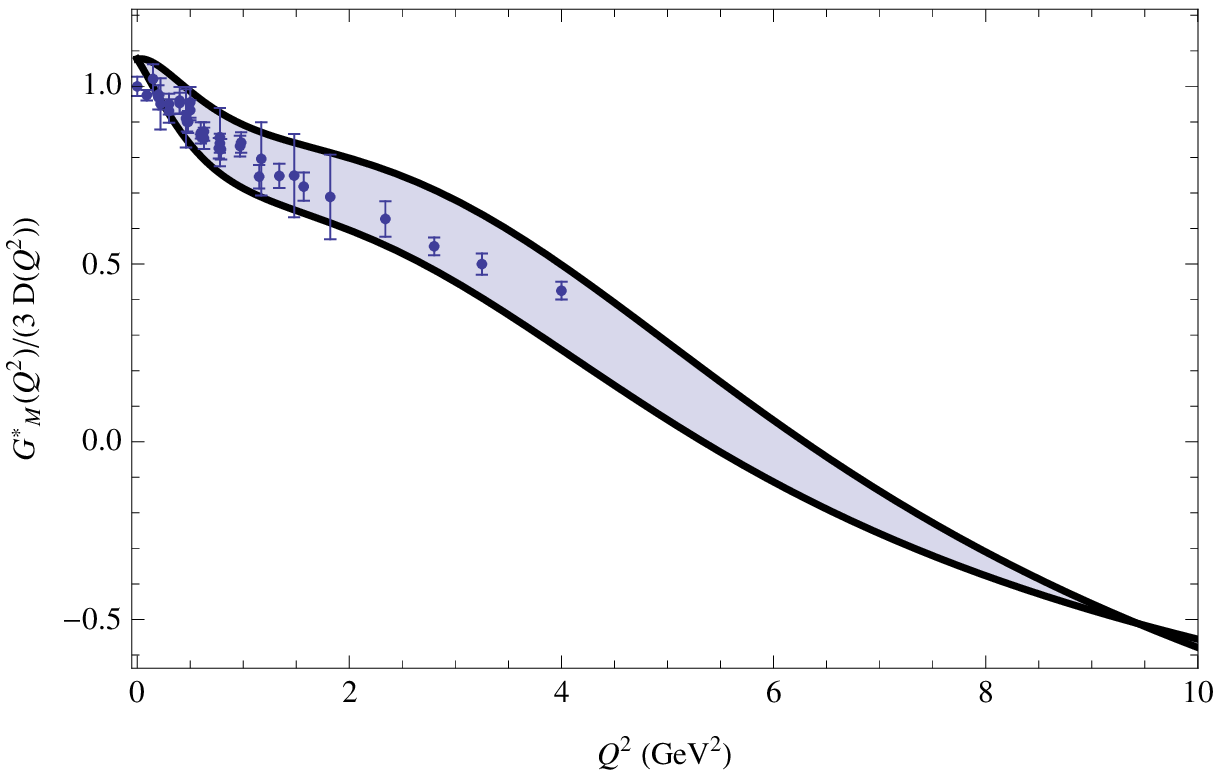,scale=.6} 
\caption{$Q^2$ dependence of the magnetic form factor 
$G_M^*(Q^2)$ divided by the dipole 
form factor $3 D(Q^2)$ up to $Q^2=$ 10~GeV$^2$. 
Our results are compared with existing data taken 
from~\cite{Bartel:1968tw,Stein:1975yy,%
Foster:1983kn,Frolov:1998pw,Kamalov:2000en,Tiator:2003xr}.                        
\label{fig12}}

\vspace*{.1cm}
\epsfig{figure=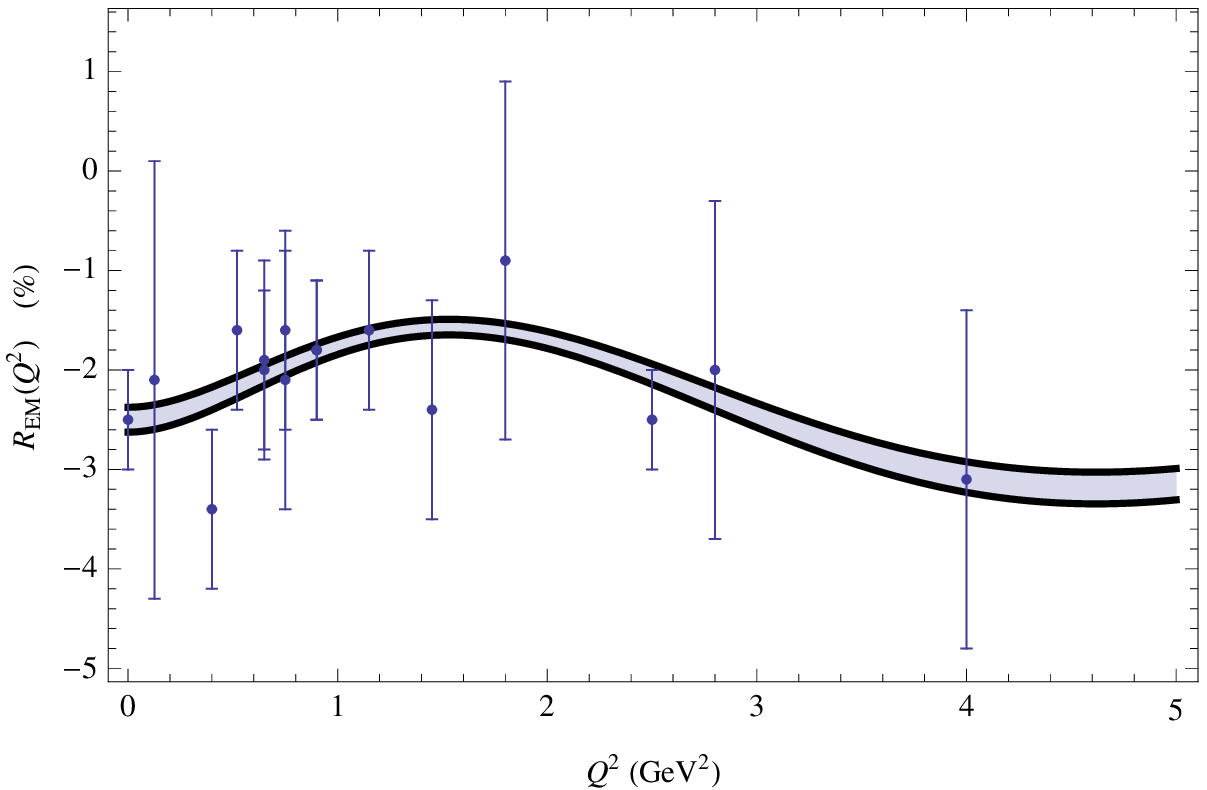,scale=.6}
\epsfig{figure=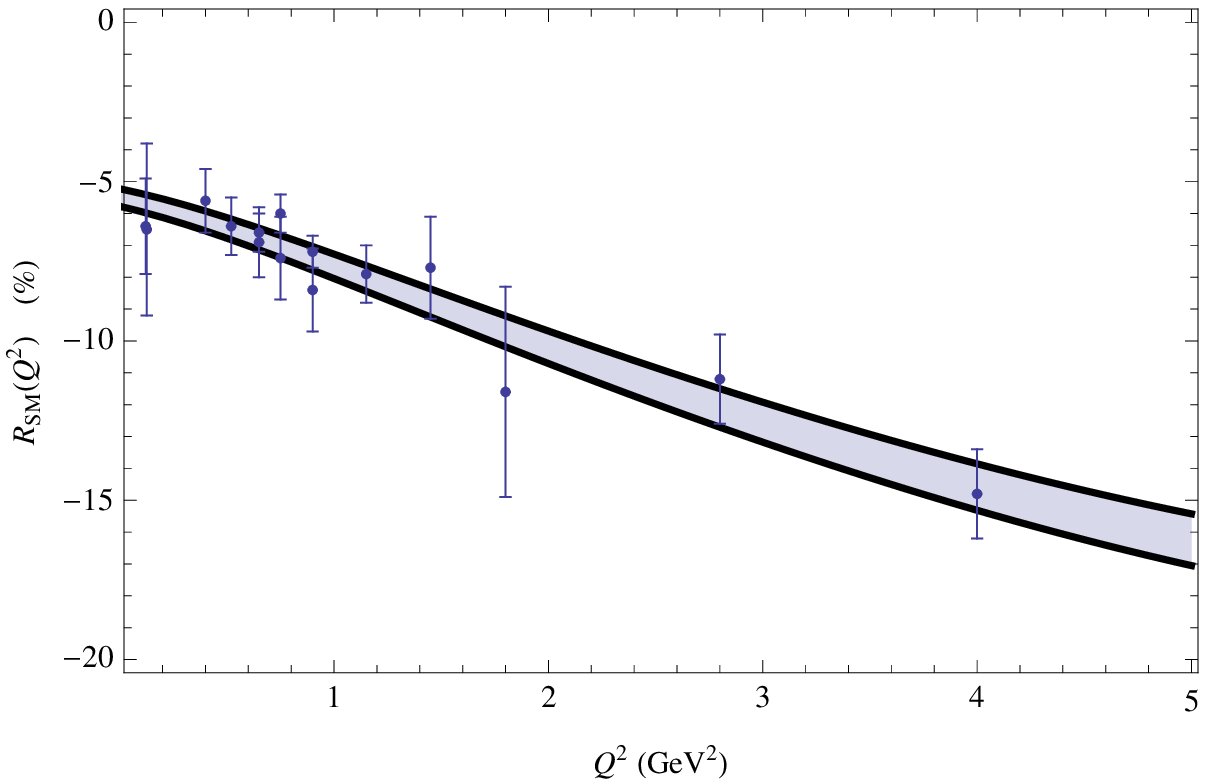,scale=.6}
\caption{$Q^2$ dependence of the ratios $R_{EM}(Q^2)$ 
(left panel) and $R_{SM}(Q^2)$ (right panel) up to 5~GeV$^2$. 
Our results are compared with data taken 
from PDG~\cite{PDG20} and 
Refs.~\cite{Beck:1997ew,Mertz:1999hp,Pospischil:2000ad}.                                                         
\label{fig13}}

\vspace*{.1cm}
\epsfig{figure=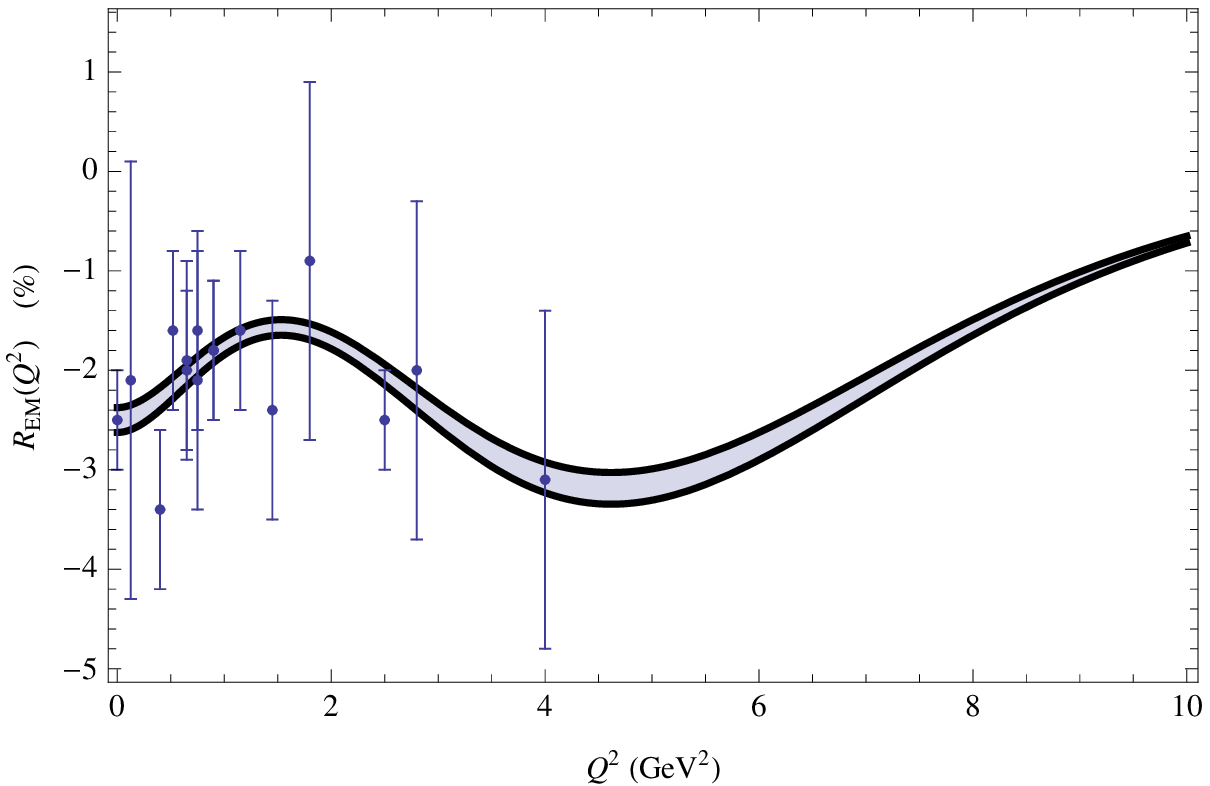,scale=.6}
\epsfig{figure=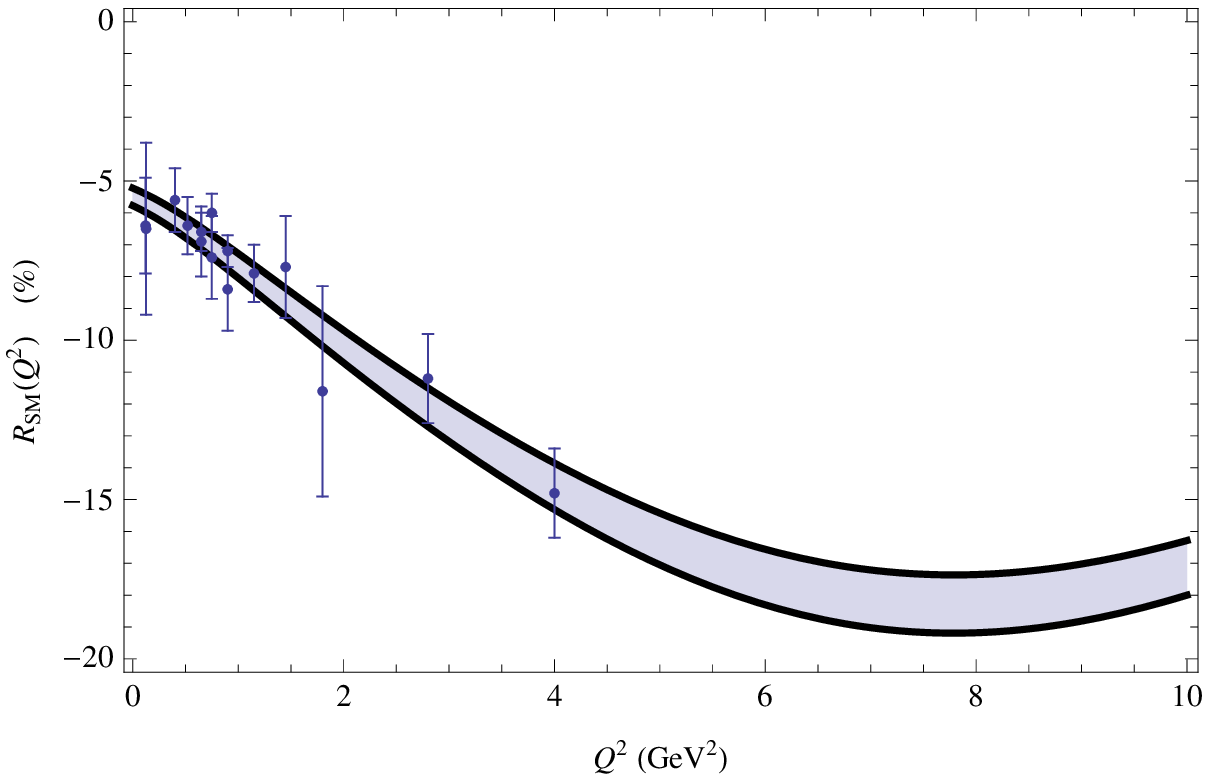,scale=.6}
\caption{$Q^2$ dependence of the ratios $R_{EM}(Q^2)$ 
(left panel) and $R_{SM}(Q^2)$ (right panel) up to 10~GeV$^2$. 
Our results are compared with data taken 
from PDG~\cite{PDG20} and 
Refs.~\cite{Beck:1997ew,Mertz:1999hp,Pospischil:2000ad}.                                                         
\label{fig14}}

\end{center}
\end{figure}

It is important to stress that at large values of $Q^2$ the form factors 
and helicity amplitudes for the electroexcitation of nucleon resonances 
are consistent with quark counting rules~\cite{Brodsky:1973kr}. 
In particular, the sets of the form factors $h_i$, $G_i$, 
($G_E$, $G_M$, $G_C$) and helicity amplitudes 
($A_{1/2}$, $A_{3/2}$, $S_{1/2}$) scale as 
\eq
& &h_i(Q^2) \sim \frac{1}{Q^{\tau+\tau^*-2}}\,, \qquad
   G_i(Q^2) \sim \frac{1}{Q^{\tau+\tau^*}}\,, \nonumber\\
& &G_E(Q^2), G_M(Q^2), G_C(Q^2) \sim \frac{1}{Q^{\tau+\tau^*-2}}\,,
\nonumber\\
& &A_{1/2}^l(Q^2), A_{3/2}^l(Q^2) \sim \frac{1}{Q^{\tau+\tau^*-1-2l}}\,,
\qquad
S_{1/2}^l(Q^2) \sim \frac{1}{Q^{\tau+\tau^*-3-2l}}\,.
\en
Model parameters (central values) 
used for each $\gamma^* N \to N^*$ transition are shown 
in Tables~\ref{tab:parameters1} and~\ref{tab:parameters2}. As in previous calculations 
we include the contributions of three leading twists.  Also, to reduce a number of 
free parameters we drop the contribution to the form factors induced by the 
couplings $g^{3M}_{\tau\tau^*}$ and $g^{4M}_{\tau\tau^*}$. 
 
\begin{table}[htb]
\begin{center}
\caption{Model parameters for $\frac{1}{2}^+ \to  \frac{1}{2}^\pm$ transitions} 
\vspace*{.1cm}

\def\arraystretch{1.25}
\begin{tabular}{|c|l|}
\hline
Transition & Choice of parameters \\ 
\hline
$N \to N(1650)$ & $c_{34} = -0.56$, $c_{45} = 0.66$, $c_{56} = 0.50$, \\
                & $c_{44} = 1.03$,  $c_{55} = 0.94$, \\ 
                &   $\eta = 1.31$,   $\zeta = -0.27$, $\xi = -0.03$ \\
\hline
$N \to N(1710)$ & $c_{33} = 0.09$, $c_{44} = 0.12$, $c_{55} = -0.05$, \\
                & $g = 1$, $\eta = -1.70$, $\lambda = 0.95$, $\zeta = 5.24$, $\xi = -8.22$ \\

\hline
$N \to \Delta(1620)$ & $c_{34} = -1.84$, $c_{45} = 2.92$, $c_{56} = -0.73$, \\
                     & $c_{44} = -1.06$, $c_{55} = 0.01$, \\
                     &    $\eta = 3.00$,  $\zeta = -0.52$, $\xi = -0.10$ \\
\hline
\end{tabular}
\label{tab:parameters1}
\end{center}
\end{table}

\begin{table}[htb]
\begin{center}
\caption{Model parameters for $\frac{1}{2}^+ \to  \frac{3}{2}^\pm, \frac{5}{2}^\pm$ transitions} 
\vspace*{.1cm}

\def\arraystretch{1.25}
\begin{tabular}{|c|l|}
\hline
Transition & Choice of parameters \\ 
\hline
$N \to N(1520)$ & $g_{34}^{1A} = -13.79$, 
                  $g_{45}^{1A} = -20.87$, 
                  $g_{56}^{1A} = - 8.70$, 
                  $g_{34}^{2B} =   1.01$, 
                  $g_{45}^{2B} = - 5.71$, 
                  $g_{56}^{2B} =   1.32$, \\
                & $g_{34}^{2C} = - 6.49$, 
                  $g_{45}^{2C} =  11.76$, 
                  $g_{56}^{2C} = - 5.04$, 
                  $g_{34}^{2D} =- 17.46$, 
                  $g_{45}^{2D} =  26.44$, 
                  $g_{56}^{2D} =- 11.01$, \\
                & $g_{34}^{1E} =- 15.97$, 
                  $g_{45}^{1E} =- 90.70$, 
                  $g_{56}^{1E} =  20.76$, 
                  $g_{34}^{1F} =- 22.27$, 
                  $g_{45}^{1F} =  40.38$, 
                  $g_{56}^{1F} =- 17.29$  \\
\hline
$N \to N(1675)$ & 
                  $g_{34}^{2A} = - 0.46$, 
                  $g_{45}^{2A} =   1.34$, 
                  $g_{56}^{2A} = - 0.77$, 
                  $g_{34}^{1B} = - 1.39$, 
                  $g_{45}^{1B} =   2.82$, 
                  $g_{56}^{1B} = - 6.60$, \\
                & $g_{34}^{1C} =  12.54$, 
                  $g_{45}^{1C} =- 51.12$, 
                  $g_{56}^{1C} =  42.51$, 
                  $g_{34}^{1D} = - 0.92$,  
                  $g_{45}^{1D} =   2.67$, 
                  $g_{56}^{1D} = - 1.53$, \\
                & $g_{34}^{2E} = - 0.70$, 
                  $g_{45}^{2E} =   1.42$, 
                  $g_{56}^{2E} = - 3.33$, 
                  $g_{34}^{2F} = - 0.55$, 
                  $g_{45}^{2F} =   2.23$, 
                  $g_{56}^{2F} = - 1.86$  \\
\hline
$N \to N(1680)$ & 
                  $g_{35}^{1A} = - 2.64$, 
                  $g_{46}^{1A} =   2.98$, 
                  $g_{57}^{1A} = - 2.13$, 
                  $g_{35}^{2B} =  11.65$, 
                  $g_{46}^{2B} =- 13.13$, 
                  $g_{57}^{2B} =   9.39$, \\
                & $g_{35}^{2C} =  34.34$, 
                  $g_{46}^{2C} =  16.31$, 
                  $g_{57}^{2C} =- 21.05$, 
                  $g_{35}^{2D} =  11.65$, 
                  $g_{46}^{2D} =- 13.13$, 
                  $g_{57}^{2D} =   9.39$, \\
                & $g_{35}^{1E} =- 17.81$, 
                  $g_{46}^{1E} =  29.05$, 
                  $g_{57}^{1E} =  11.36$, 
                  $g_{35}^{1F} = 211.90$, 
                  $g_{46}^{1F} = 100.61$,  
                  $g_{57}^{1F} = 129.91$  \\
\hline
$N \to N(1700)$ & $g_{34}^{1A} =   0.18$, 
                  $g_{45}^{1A} = - 0.28$, 
                  $g_{56}^{1A} =   0.16$, 
                  $g_{34}^{2B} = - 0.09$, 
                  $g_{45}^{2B} =   0.17$, 
                  $g_{56}^{2B} =   0.12$, \\
                & $g_{34}^{2C} =   0.02$, 
                  $g_{45}^{2C} =   0.06$, 
                  $g_{56}^{2C} = - 0.08$, 
                  $g_{34}^{2D} = - 0.20$, 
                  $g_{45}^{2D} =   0.31$, 
                  $g_{56}^{2D} = - 0.17$, \\
                & $g_{34}^{1E} =   1.48$, 
                  $g_{45}^{1E} = - 2.95$, 
                  $g_{56}^{1E} = - 2.03$, 
                  $g_{34}^{1F} = - 0.30$, 
                  $g_{45}^{1F} = - 1.06$, 
                  $g_{56}^{1F} =   1.46$  \\
\hline
$N \to N(1720)$ & 
                  $g_{35}^{2A} = - 11.58$, 
                  $g_{46}^{2A} =   34.50$, 
                  $g_{57}^{2A} = - 22.60$, 
                  $g_{35}^{1B} =   79.15$, 
                  $g_{46}^{1B} =   67.11$, 
                  $g_{57}^{1B} =   29.95$, \\
                & $g_{35}^{1C} = - 36.53$, 
                  $g_{46}^{1C} =  105.65$, 
                  $g_{57}^{1C} = - 58.59$, 
                  $g_{35}^{1D} =    0.16$, 
                  $g_{46}^{1D} = -  0.47$, 
                  $g_{57}^{1D} =    0.31$, \\
                & $g_{35}^{2E} =   12.14$, 
                  $g_{46}^{2E} =   10.29$, 
                  $g_{57}^{2E} =    4.59$, 
                  $g_{35}^{2F} =    5.56$, 
                  $g_{46}^{2F} = - 16.07$,  
                  $g_{57}^{2F} =    8.91$  \\
\hline
$N \to N'(1720)$ & 
                  $g_{35}^{2A} =   30.18$, 
                  $g_{46}^{2A} = - 54.02$, 
                  $g_{57}^{2A} =   24.82$, 
                  $g_{35}^{1B} = -  1.15$, 
                  $g_{46}^{1B} =    6.41$, 
                  $g_{57}^{1B} = -  2.87$, \\
                & $g_{35}^{1C} = - 17.77$, 
                  $g_{46}^{1C} =  107.19$, 
                  $g_{57}^{1C} = - 62.71$, 
                  $g_{35}^{1D} =    8.48$, 
                  $g_{46}^{1D} = - 15.18$, 
                  $g_{57}^{1D} =    6.97$, \\
                & $g_{35}^{2E} = -  0.60$, 
                  $g_{46}^{2E} =    3.37$, 
                  $g_{57}^{2E} = -  1.51$, 
                  $g_{35}^{2F} = -  3.57$, 
                  $g_{46}^{2F} =   21.56$,  
                  $g_{57}^{2F} = - 12.61$  \\
\hline
$N \to \Delta(1232)$ & 
                  $g_{35}^{2A} = - 14.00$, 
                  $g_{46}^{2A} =   22.38$, 
                  $g_{57}^{2A} = - 10.13$, 
                  $g_{35}^{1B} = -  0.26$, 
                  $g_{46}^{1B} =    0.18$, 
                  $g_{57}^{1B} =    2.05$, \\
                & $g_{35}^{1C} =    0.97$, 
                  $g_{46}^{1C} = -  3.01$, 
                  $g_{57}^{1C} =    2.05$, 
                  $g_{35}^{1D} = -  6.49$, 
                  $g_{46}^{1D} =   10.39$, 
                  $g_{57}^{1D} = -  4.70$, \\
                & $g_{35}^{2E} = -  3.82$, 
                  $g_{46}^{2E} =    0.26$, 
                  $g_{57}^{2E} =    2.97$, 
                  $g_{35}^{2F} =    0.57$, 
                  $g_{46}^{2F} = -  1.77$,  
                  $g_{57}^{2F} =    1.20$  \\
\hline
$N \to \Delta(1700)$ & 
                  $g_{34}^{1A} = - 0.01$, 
                  $g_{45}^{1A} = - 0.59$, 
                  $g_{56}^{1A} =   0.40$, 
                  $g_{34}^{2B} = - 0.53$, 
                  $g_{45}^{2B} =   0.70$, 
                  $g_{56}^{2B} = - 0.27$, \\
                & $g_{34}^{2C} =   1.02$, 
                  $g_{45}^{2C} =   2.70$, 
                  $g_{56}^{2C} = - 0.97$, 
                  $g_{34}^{2D} = - 0.05$,  
                  $g_{45}^{2D} =   3.57$, 
                  $g_{56}^{2D} = - 2.44$, \\
                & $g_{34}^{1E} = - 0.93$, 
                  $g_{45}^{1E} =   1.23$, 
                  $g_{56}^{1E} = - 0.47$, 
                  $g_{34}^{1F} =  12.09$, 
                  $g_{45}^{1F} =  32.11$, 
                  $g_{56}^{1F} = - 11.49$  \\
\hline
\end{tabular}
\label{tab:parameters2}
\end{center}
\end{table}

Our results for the $Q^2$ dependence of the helicity amplitudes in the
$\gamma^* N \to N^*$ transitions including a variation of 
the parameters (up to 20\%)  are fully displayed in Figs.~\ref{fig1}-\ref{fig11}. 
In Figs.~\ref{fig1}-\ref{fig3} we present the results for the modes with 
nucleon resonances having spin $\frac{1}{2}$, which were not considered by us before and in addition 
in Figs.~\ref{fig4}-\ref{fig14} we display the results for the nucleon resonances 
with higher spins $\frac{3}{2}$ and $\frac{5}{2}$. 
We compare our reuslts to data from the CLAS Collaboration 
(JLab)~\cite{Aznauryan:2009mx,Dugger:2009pn,Aznauryan:2012ec,Mokeev:2012vsa,
Mokeev:2013kka,Park:2014yea,Mokeev:2015lda,Golovatch:2018hjk},  
other experiments~\cite{Bartel:1968tw,Stein:1975yy,Foster:1983kn,Beck:1997ew,%
Frolov:1998pw,Mertz:1999hp,Pospischil:2000ad,Blanpied:2001ae}, 
and world data analyses~\cite{Burkert:2002zz,Kamalov:2000en,Tiator:2003xr}.                         
Also we consider in detail the observables of the $\gamma^* N \to \Delta(1232)$ 
transitions: helicity amplitudes (Fig.~\ref{fig11}), 
the $Q^2$ dependence of the magnetic form factor 
$G_M^*(Q^2)$ divided by the dipole 
form factor $3 D(Q^2)$ (Fig.~\ref{fig12}),  where $D(Q^2)=1/(1+Q^2/0.71 \, {\rm GeV}^2)^2$, 
the $Q^2$ dependence of the $R_{EM} = E/M$ and $R_{SM} = S/M$ ratios 
(Fig.~\ref{fig13} up to 5 GeV$^2$ and Fig.~\ref{fig14} up to 10 GeV$^2$)  
magnetic dipole $\mu_{N\Delta}$ and electric quadrupole $Q_{N\Delta}$ 
moments: 
\eq 
R_{EM}(Q^2) &=& \frac{A_{1/2}(Q^2) - A_{3/2}(Q^2)/\sqrt{3}} 
{A_{1/2}(Q^2) + A_{3/2}(Q^2) \sqrt{3}} \,, \nonumber\\
R_{SM}(Q^2) &=& \frac{S_{1/2}(Q^2) \sqrt{2}} 
{A_{1/2}(Q^2) + A_{3/2}(Q^2) \sqrt{3}} \,, \\
\mu_{N\Delta} &=& \sqrt{\frac{M_\Delta}{M_N}} \, G_M^*(0) 
\,, \nonumber\\
Q_{N\Delta} &=& - \frac{6}{M_N E_N} \, \sqrt{\frac{M_\Delta}{M_N}} \, G_E^*(0) 
\,. \nonumber\\
\en   
Note the magnetic $G_M^*(Q^2)$ and electric $G_E^*(Q^2)$ 
form factors are normalized as~\cite{Tiator:2003xr}: 
\eq 
G_M^*(Q^2) &=& G_M(Q^2) \, \frac{c_\Delta}{F_1^+} = 
- c_\Delta \, \Big(A_{1/2}(Q^2) + A_{3/2}(Q^2) \sqrt{3}\Big) 
\,, \nonumber\\ 
G_E^*(Q^2) &=& G_E(Q^2) \, \frac{c_\Delta}{F_1^+} = 
c_\Delta \, \Big(A_{1/2}(Q^2) - \frac{A_{3/2}(Q^2)}{\sqrt{3}}\Big) 
\,, \nonumber\\ 
\en 
where 
\eq 
c_\Delta = \frac{M_N}{|{\bf p}|} \, \sqrt{\frac{M_N E_N}{4 \pi \alpha M_\Delta}} \,. 
\en 
In Table~\ref{tab:REM} our results for $R_{EM}(0)$ are compared 
with existing data (PDG~\cite{PDG20}, MAMI experiment~\cite{Beck:1997ew}, 
LEGS Collaboration~\cite{Blanpied:2001ae}) and some theoretical approaches 
[model bases on partial-Wave analysis (SAID)~\cite{Arndt:2002xv}, 
approach based on dispersion relations and unitarity (DR)~\cite{Hanstein:1997tp} and 
relativistic quark model (RQM)~\cite{Faessler:2006ky}].  

For $R_{SM}(0)$ we get $- 5.5 \pm 0.5$. 
Our predictions for 
the moments $\mu_{N\Delta}$ and $Q_{N\Delta}$ 
\eq 
\mu_{N\Delta} = 3.7\pm 0.4 \,, \qquad 
Q_{N\Delta} = - (0.09 \pm 0.01) \, {\rm fm}^2\,. 
\en  
are in good agreement with data (LEGS Collaboration~\cite{Blanpied:2001ae}):  
\eq 
\mu_{N\Delta} = 3.642 \pm 0.019 \pm 0.085 \,, \qquad 
Q_{N\Delta} = - (0.108 \pm 0.009 \pm 0.034) \, {\rm fm}^2\,. 
\en 
with the Mainz 
multipole analysis~\cite{Tiator:2003xr}: 
\eq 
\mu_{N\Delta} = 3.46 \pm 0.03 \,, \qquad 
Q_{N\Delta} = - (0.0846 \pm 0.0033) \, {\rm fm}^2\,. 
\en  

\begin{table}[ht]
\begin{center}
\caption{ $R_{EM}(0)$ ratio in (\%) }
\vspace*{.1cm}

\def\arraystretch{1.25}
    \begin{tabular}{l}
\hline 
$-2.5 \pm 0.5$            \hfill PDG~\cite{PDG20} \\ 
\hline 
$-2.5 \pm 0.1 \pm 0.2$    \hfill MAMI~\cite{Beck:1997ew} \\ 
\hline 
$-3.07 \pm 0.26 \pm 0.24$ \hfill LEGS~\cite{Blanpied:2001ae} \\ 
\hline 
$-2.0  \pm 0.2$           \hfill SAID~\cite{Arndt:2002xv} \\ 
\hline 
$ -2.54 \pm 0.10$         \hfill DR~\cite{Hanstein:1997tp}   \\
\hline 
$ -3.02 \pm 0.08$         \hfill RQM~\cite{Faessler:2006ky}   \\
\hline 
$-2.5 \pm 0.5$ \hfill (Our results) \\
\hline
\label{tab:REM}
\end{tabular}
\end{center}
\end{table}

\section{Summary}

We extended our formalism based on a soft-wall AdS/QCD approach to the 
description of the electro-couplings of nucleons with nucleon resonances 
with high spins. All form factors and helicity amplitudes 
characterizing the electromagnetic transitions between nucleons 
and nucleon resonances are consistent with quark counting 
rules~\cite{Brodsky:1973kr}. We fix free parameters in our approach using 
data from the CLAS Collaboration~\cite{Aznauryan:2009mx,Dugger:2009pn,%
Aznauryan:2012ec,Mokeev:2012vsa,Mokeev:2013kka,Park:2014yea,%
Mokeev:2015lda,Golovatch:2018hjk} and 
a compilation of the world analyses of the $N\pi$ electroproduction 
data~\cite{Burkert:2002zz}. In our calculations we adopt a variation 
of free parameters up to 20\%. The main success of our approach is based on 
analytical implementation of quark counting rules~\cite{Brodsky:1973kr}. 

\begin{acknowledgments}

This work was funded by ``Verbundprojekt 05P2018 - Ausbau von ALICE                                  
am LHC: Jets und partonische Struktur von Kernen'' (F\"orderkennzeichen: 05P18VTCA1),  
``Verbundprojekt 05A2017 - CRESST-XENON: Direkte Suche nach Dunkler                              
Materie mit XENON1T/nT und CRESST-III. Teilprojekt 1''
(F\"orderkennzeichen 05A17VTA)'', by ANID (Chile) under
Grant No. 7912010025,  by ANID PIA/APOYO AFB180002,
by FONDECYT (Chile) under Grants No. 1191103 and No. 1180232, 
by the Tomsk State University Competitiveness Enhancement Program 
``Research of Modern Problems of Quantum Field Theory and Condensed Matter Physics''
and Tomsk Polytechnic University Competitiveness Enhancement Program (Russia). 

\end{acknowledgments} 

\appendix 

\section{AdS/QCD action for description of the $\gamma + \frac{1}{2}^+  \to \frac{1}{2}^\pm$ 
transitions} 
\label{action12}

The AdS/QCD action for description of the 
$\gamma + \frac{1}{2}^+  \to \frac{1}{2}^\pm$                              
transitions contains a free part $S_0$, describing the
confined dynamics of AdS fields, and interaction part $S_{\rm int}$,
describing interactions of fermions with vector field
with
\eq\label{actionS12}
S   &=& S_0 + S_{\rm int}\,, \nonumber\\[3mm]
S_0 &=& \int d^4x dz \, \sqrt{g} \, e^{-\varphi(z)} \,
\biggl\{ {\cal L}_N(x,z) + {\cal L}_{N^*}(x,z)
+ {\cal L}_V(x,z)
\biggr\} \,, \nonumber\\[3mm]
S_{\rm int} &=& \int d^4x dz \, \sqrt{g} \, e^{-\varphi(z)} \,
\biggl\{
{\cal L}_{VNN}(x,z) + {\cal L}_{VN^*N}(x,z) + {\cal L}_{VN^*N^*}(x,z)
\biggr\} \,, 
\en
where ${\cal L}_N$, ${\cal L}_{N^*}$, ${\cal L}_V(x,z)$ and
${\cal L}_{VNN}(x,z)$, ${\cal L}_{VN^*N}(x,z)$, ${\cal L}_{VN^*N^*}(x,z)$ 
are the free and interaction Lagrangians, respectively. 
See details in Refs.~\cite{Gutsche:2017lyu,%
Gutsche:2019jzh,Gutsche:2019yoo}, term ${\cal L}_V(x,z)$ 
is specified in Eq.~(\ref{actionS2}). Below we specify the interaction 
Lagrangian ${\cal L}_{VN^*N}(x,z)$, relevant for the 
$\gamma + \frac{1}{2}^+  \to \frac{1}{2}^\pm$ transitions. 
In particular,
\eq
{\cal L}_{VN^*N}(x,z) &=&
\sum\limits_{i=+,-; \,\tau\tau^*} \, 
\bar\psi_{i,\tau^*}^{N^*}(x,z) \, \hat{\cal V}^{N^*N}_{i,\tau\tau^*}(x,z) \,
\psi_{i,\tau}^N(x,z) \, + \, {\rm H.c.}\,,
\en 
where 
\eq
\hat{\cal V}^{N^*N}_{\pm,\tau\tau^*}(x,z)  &=&  
c_{\tau\tau^*} \, Q \, \Gamma^M  V_M(x,z) \,+\, 
d_{\tau\tau^*} \, \biggl[\pm \,
\frac{i}{4} \, \eta \,  [\Gamma^M, \Gamma^N] \, V_{MN}(x,z)
\nonumber\\[3mm] 
&\pm& \frac{i}{4} \, \lambda  \, z^2 \,
[\Gamma^M, \Gamma^N] \, \partial^K\partial_KV_{MN}(x,z)
\, \pm \,  g \, \Gamma^M \, i\Gamma^z \, V_M(x,z)
\nonumber\\[3mm]
&+& \zeta \, z \, \Gamma^M  \, \partial^N V_{MN}(x,z)
\, \pm \, \xi \, z \, \Gamma^M \, i\Gamma^z \, \partial^N V_{MN}(x,z) \biggr]
\,. 
\en
Here $c_{\tau\tau^*}$, $d_{\tau\tau^*}$, 
$\eta$, $\lambda$, $g$, $\zeta$, and $\xi$ are 
the couplings fixed from description of data on the $Q^2$ dependence 
of the $\gamma + \frac{1}{2}^+  \to \frac{1}{2}^\pm$ transitions. 
In case of the $\gamma + \frac{1}{2}^+  \to \frac{1}{2}^+$ transitions we 
use $c_{\tau\tau^*} \equiv d_{\tau\tau^*}$.

\end{document}